\documentclass[journal,onecolumn]{IEEEtran}
\usepackage{cite}
\usepackage{amsmath,amssymb,amsfonts}
\usepackage{algorithmic}
\usepackage[ruled,vlined,linesnumbered]{algorithm2e}
\usepackage{graphicx}
\usepackage{caption}
\usepackage{subcaption}
\usepackage{paralist}
\usepackage{textcomp}
\usepackage{xcolor}
\usepackage{cite}
\usepackage{graphicx}
\usepackage{amsfonts}
\usepackage{amssymb}
\usepackage{amsbsy}
\usepackage{amsthm}
\usepackage[mathscr]{eucal}
\usepackage{comment}
\usepackage{paralist}
\usepackage{boxedminipage}
\usepackage{array}
\usepackage{multirow}
\usepackage{float}
\usepackage{color}
\usepackage{mathtools}
\usepackage{bm}

\ifCLASSINFOpdf
\else
\fi
\begin{document}
\begin{center}
 Reconfigurable Intelligent Surface-Aided Wireless Power Transfer Systems: Analysis and Implementation

\vspace{10cm}
\textcolor{blue}{This work has been submitted to the IEEE for possible publication. Copyright may be transferred without notice, after which this version may no longer be accessible.}
\end{center}
\newpage
%
\title{Reconfigurable Intelligent Surface-Aided Wireless Power Transfer Systems: Analysis and Implementation}
%
%
%

\author{Nguyen~Minh~Tran,~\IEEEmembership{Student Member,~IEEE,}
Muhammad~Miftahul~Amri,~\IEEEmembership{Student Member,~IEEE,}
Je~Hyeon~Park,~\IEEEmembership{Student Member,~IEEE,}
Dong~In~Kim,~\IEEEmembership{Fellow,~IEEE,}
and~Kae~Won~Choi,~\IEEEmembership{Senior Member,~IEEE}
\thanks{\textcolor{blue}{This work has been submitted to the IEEE for possible publication. Copyright may be transferred without notice, after which this version may no longer be accessible.}
}}

\maketitle

\begin{abstract}
Reconfigurable intelligent surface (RIS) is a promising technology for RF wireless power transfer (WPT) as it is capable of beamforming and beam focusing without using active and power-hungry components. In this paper, we propose a multi-tile RIS beam scanning (MTBS) algorithm for powering up internet-of-things (IoT) devices. Considering the hardware limitations of the IoT devices, the proposed algorithm requires only power information to enable the beam focusing capability of the RIS. Specifically, we first divide the RIS into smaller RIS tiles. Then, all RIS tiles and the phased array transmitter are iteratively scanned and optimized to maximize the receive power. We elaborately analyze the proposed algorithm and build a simulator to verify it. Furthermore, we have built a real-life testbed of RIS-aided WPT systems to validate the algorithm. The experimental results show that the proposed MTBS algorithm can properly control the transmission phase of the transmitter and the reflection phase of the RIS to focus the power at the receiver. Consequently, after executing the algorithm, about 20 dB improvement of the receive power is achieved compared to the case that all unit cells of the RIS are in OFF state. By experiments, we confirm that the RIS with the MTBS algorithm can greatly enhance the power transfer efficiency.

\end{abstract}

\begin{IEEEkeywords}
Reconfigurable intelligent surface (RIS), RF wireless power transfer (WPT), beam focusing, energy harvesting efficiency.
\end{IEEEkeywords}

%
\IEEEpeerreviewmaketitle

\section{Introduction} \label{Introduction}

\IEEEPARstart{I}{N} the near future, a massive number of internet-of-things (IoT) devices are expected to be deployed \cite{ref1}. Powering these billions of IoT devices is challenging. A surging maintenance cost makes the traditional methods with power cords and removable batteries not suitable anymore~\cite{ref2}. Recently, the radio frequency (RF) wireless power transfer (WPT) is emerging as a potential technology for resolving this difficulty~\cite{ref3,ref4}. In contrast to inductive or magnetic resonance coupling, RF WPT has the advantage of long charging distance. However, its transfer efficiency degrades quickly as the distance increases. Recently, much higher efficiency can be achieved with RF WPT by beamforming technology. Utilizing the phased array antenna (PAA), one can perform beamforming to focus the electromagnetic (EM) wave at the receiver. Nonetheless, in the PAA system, each radiating element should be equipped with various RF components such as amplifiers, phase shifters, and attenuators. This results in very high complexity, high implementation cost, and higher power consumption in the system, especially in a large-scale system.

Recently, reconfigurable intelligent surface (RIS) is emerged as a promising technology for RF WPT since it is capable of beamforming and beam focusing without using active components. The RIS is also known as other alternative names such as intelligent reflecting surface (IRS) \cite{ref5}, large intelligent surface (LIS) \cite{ref6}, software-controlled metasurface \cite{ref7}, or programmable coding metasurface \cite{ref8,ref9}. An RIS consists of hundreds to thousands of passive reflecting unit cells with a special design structure. With a control element (e.g., a PIN diode) integrated, one can electronically control and change the characteristics (e.g., phase, magnitude) of the incoming wave upon each unit cell in the RIS. By intelligently controlling the reflection phase of each unit cell in the RIS, power beams can be focused to the desired positions. 

\begin{figure}[!htb]
    \centering
    \includegraphics[trim =8cm 5.25cm 1cm 5cm,clip =true, width = 0.5\textwidth]{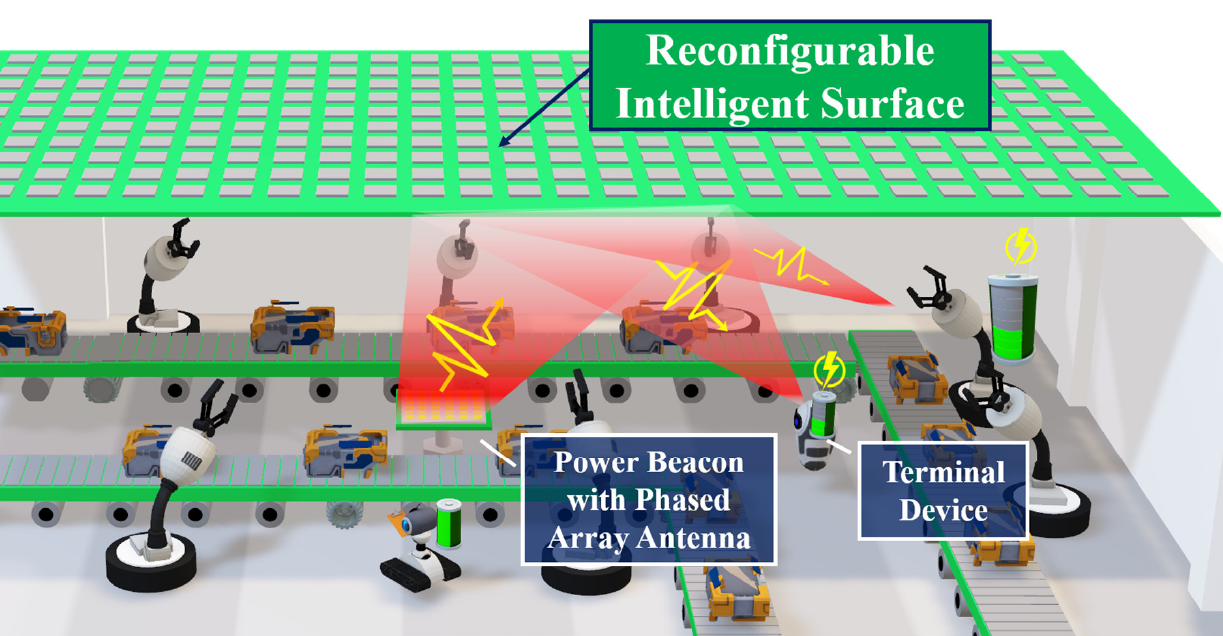}
    \caption{RIS-aided WPT system conceptual application scenario.}
    \label{fig:RISconceptapp}
\end{figure}

The RIS enables beamforming and beam focusing without using active and power-hungry components. In other words, the RIS ensures lower loss in RF wireless energy transfer as compared to the existing technologies only using PAA. Additionally, RISs can be massively manufactured at a very low cost. Then, one can easily deploy a very large scale of RISs in the walls of a building or a room to boost up the power transfer efficiency. These features make it preferable for RF wireless power transfer applications. One potential application scenario of the RIS-aided WPT system is given in Fig.~\ref{fig:RISconceptapp}. In this scenario, the RIS is installed in the ceiling of a smart-automated factory for assisting the WPT system. The RIS reflects the EM beam from a power beacon, then it focuses the reflected wave at the devices on the ground. Accordingly, the RIS helps the WPT system to enhance the power transfer efficiency and extend the power transfer range.

Although the RIS offers considerable benefits compared to conventional PAA systems, several challenges need to be addressed when it comes to practice. The channel estimation is one of the challenging tasks that needs to be tackled for realizing the RIS-aided WPT. To perform beamforming and beam focusing, one should estimate the channel between the transmitter, receiver, and the RIS. 

Some recent works have proposed effective channel training strategies for RIS channel estimation. The authors of \cite{ref10} have proposed a channel estimation protocol that turns on each single element while keeping others off. This approach is practically difficult to realize for the following reasons. The cost and complexity of the system increase if we control both the phase and magnitude of the unit cells separately. Moreover, the reflection from only one unit cell is not noticeable since it is submerged in the other strong signals. Similar channel training method is introduced in \cite{ref11,ref12}. The authors of \cite{ref13} proposed a more practical training method called DFT-Hadamard-based basis training reflection matrix. The authors considered practical discrete phase shifts and demonstrated that the proposed method operates effectively. Another efficient channel estimation protocol is proposed in \cite{ref14} for orthogonal frequency division multiplexing (OFDM) systems. A pre-design RIS reflection pattern is obtained by using a two-dimensional discrete Fourier transform (2D-DFT) matrix. This work is further extended in \cite{ref15,ref16}.
 
The potential use of RIS in RF WPT systems has been analyzed and demonstrated in \cite{ref17,ref18,ref19,ref20,ref21,ref22}. The work~\cite{ref17} considered and analyzed an RIS-aided simultaneous wireless information and power transfer (SWIPT) system. The authors jointly optimized the transmit precoding matrices of the base station and the passive phase shift matrix of the RIS. The results show that employing RISs in SWIPT greatly enhances the system performance. Another work on using RIS for WPT is proposed in \cite{ref18}. The authors maximize the receive power by jointly optimizing the beamformer at the transmitter and the phase shifts at the RIS. However, this work assumes perfect channel state information (CSI) for simplifying the problem. Similar work has been done in \cite{ref19} with perfect channel estimation assumption. 
Practical beamforming for the RIS is demonstrated in \cite{ref21,ref22} by experiments. 
In our previous work \cite{ref21}, the experimental results with real-life testbeds have proven that RIS can greatly enhance the power transfer efficiency.
The work \cite{ref21} has reported at most 20 dB enhancement in signal-to-noise radio (SNR) by RIS for the wireless communications systems.
 
The above-mentioned works are done under the strong assumption that both the phase and magnitude of the received signal are known at the receiver. However, in practice, it is very hard to obtain the accurate phase information at the receiver because of the frequency drift and phase noise of the carrier frequency sources. Due to a low switching speed of RIS unit cells, stabilizing the phase of the carrier frequency source within one RIS state or tracking it over multiple RIS states is almost impossible. Then, the channel estimation should be done only based on the magnitude or power of the receive signal. One may come up with a typical beam scanning method with a known codebook for the entire RIS. This method may work if the receiver is in the far-field region of the RIS. However, with a typical scanning method, the energy transfer efficiency dramatically deteriorates in the radiative near-field region, which is a common situation in the large-scale RIS system. We have to come up with a solution for enabling the beam focusing capability of the RIS to handle this situation.
 
Therefore, in this paper, we propose a multi-tile RIS beam scanning (MTBS) algorithm to overcome this challenge. We divide the whole RIS into smaller RIS tiles to ensure that the transmitter and receiver are in the far-field region of each individual RIS tile. Moreover, instead of controlling each unit cell separately, we introduce high-level direction control and phase control parameters to efficiently control an RIS tile. Subsequently, based only on the receive power level, we iteratively optimize the control parameters of the RIS tiles and transmitter to maximize the receive power. We have elaborately analyzed the proposed model and derived a closed-form expression for the receive power with respect to the control parameters. The MTBS algorithm is designed based on the findings from the analysis.
 
We have built a real-life testbed of the RIS-aided WPT and performed experiments to verify the effectiveness of the proposed method. The prototype system consists of a transmitter, a receiver, and an RIS. The RIS comprises 16$\times$16 one-bit guided-wave unit cells. The transmitter is a phased array antenna that is capable of beam steering. Multiple rectennas are incorporated into the receiver to efficiently harvest the EM wave power. By experiments, we have shown that the MTBS algorithm works well with different RIS tile sizes in the test scenarios. Almost 20 dB gain in the receive power can be achieved by the MTBS algorithm compared to the case that all unit cells of the RIS are in OFF state. Furthermore, we have measured the power transfer efficiency according to the distance between the transmitter and receiver.
To the best of our knowledge, the analysis and experiments on RIS-aided WPT systems with multi-tile RIS beam scanning method have not been done in any previous work.

In summary, the contributions of this paper are threefold.
\begin{itemize}
    \item We elaborately analyze the multi-tile RIS-aided WPT scheme. Specifically, we divide the RIS into smaller RIS tiles. Moreover, we suggest high-level direction and phase control parameters to handle each RIS tile and the transmitter. Consequently, the closed-form formula for the receive power with respect to the control parameters has been derived.
    \item We propose a multi-tile RIS beam scanning (MTBS) algorithm for enhancing the performance of the RIS-aided WPT system. With only receive power information at the transmitter, the proposed algorithm can realize the beam focusing capability of the RIS.
    \item A real-life testbed of the RIS-aided WPT system is built to verify the proposed algorithm. We have shown experimental results to demonstrate the effectiveness of the MTBS algorithm. No previous works have conducted such comprehensive experiments.
\end{itemize}

The rest of the paper is organized as follows. Section~\ref{section:system model} presents the overall multi-tile RIS-aided RF-WPT system model and the control model of the RIS tile and transmitter. In Section~\ref{section:WPT}, we provide an intuitive analysis on the end-to-end WPT performance given the beam steering control parameters. 
The multi-tile RIS beam scanning algorithm is presented in Section~\ref{section:powerscanning}. The RIS design, system implementation, and experimental setups are given in Section~\ref{sec:implementation}. Simulation and experimental results are presented in Section~\ref{Sec:Results}, and the paper is concluded with Section~\ref{section:conclusion}.

\section{System Model}\label{section:system model}

\subsection{RIS-Aided Wireless Power Transfer System Spatial Model}

\begin{figure}[!htb]
    \centering
    \includegraphics[trim = 6cm 5cm 4cm 4.5cm, clip =true,width = 0.48\textwidth]{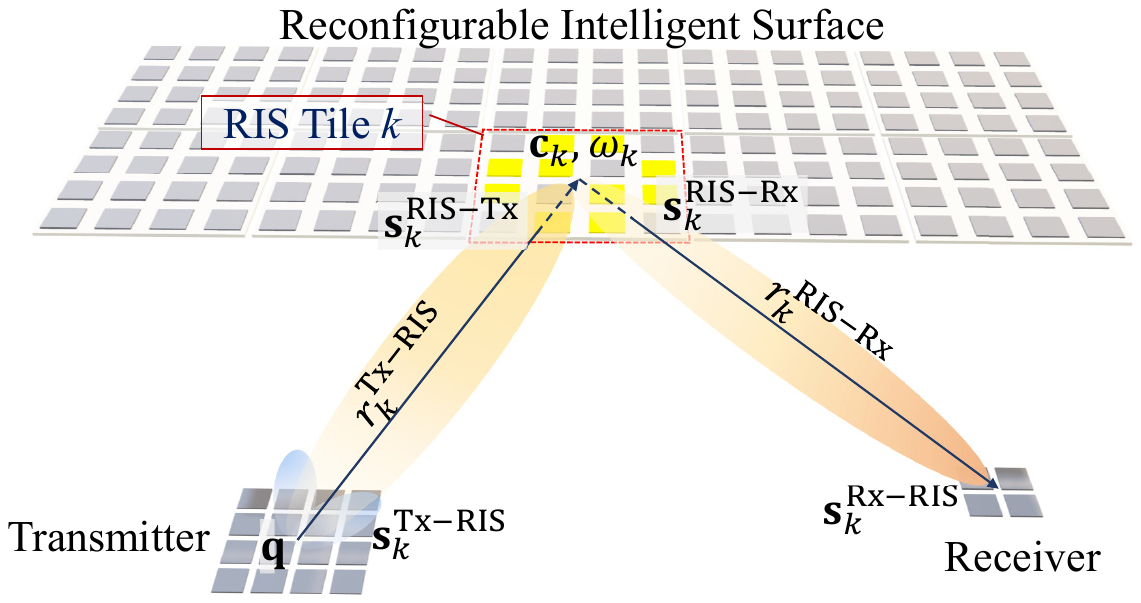}
    \caption{RIS-aided WPT system.}
    \label{fig:sysmod}
\end{figure}

The RF-WPT system in consideration consists of one multi-antenna transmitter, one multi-antenna receiver, and $K$ RIS tiles as presented in Fig.~\ref{fig:sysmod}.
The EM wave conveying the wireless power is transmitted from the transmitter, reflected by the RIS tiles, and received by the receiver.
The frequency of the EM wave is denoted by $f$, and the free-space wavelength of the EM wave is denoted by $\lambda$.

The transmitter and receiver are rectangular planar antenna arrays with $M^\text{Tx}\times N^\text{Tx}$ antenna elements and $M^\text{Rx}\times N^\text{Rx}$ antenna elements, respectively.
The transmitter and receiver have their own local coordinate system, in which the antenna elements are placed along the x-axis and y-axis on the x-y plane.
Each antenna element in the transmitter or receiver is indexed by a tuple $(m,n)$.
The antenna spacing between adjacent antenna elements in the transmitter is given by $\delta^\text{Tx,x}$ and $\delta^\text{Tx,y}$ in x and y directions, respectively.
The antenna spacing is denoted by $\delta^\text{Rx,x}$ and $\delta^\text{Rx,y}$ for the receiver.
The position of antenna element $(m,n)$ of the transmitter in the x-y plane of the local coordinate system is 
\begin{align}
{\bold u}^\text{Tx}_{m,n} = (\delta^\text{Tx,x}\kappa_m^{M^\text{Tx}}, \delta^\text{Tx,y}\kappa_n^{N^\text{Tx}})^T
\end{align}
where $\kappa_j^J$ is the $j$th grid point in the uniform grid with size $J$ centered at the origin, that is
\begin{align}\label{eq:grid}
    \kappa_j^J = j-J/2-1/2.
\end{align}
Similarly, the position of antenna element $(m,n)$ of the receiver is given by
\begin{align}
{\bold u}^\text{Rx}_{m,n} = (\delta^\text{Rx,x}\kappa_m^{M^\text{Rx}}, \delta^\text{Rx,y}\kappa_n^{N^\text{Rx}})^T.
\end{align}

In the RF-WPT system, there are $K$ rectangular-shaped RIS tiles, each of which is indexed by $k=1,\ldots,K$.
RIS tile $k$ has $M^\text{RIS}_k\times N^\text{RIS}_k$ unit cells.
In the local coordinate system for each RIS tile, the unit cells are placed along the x-axis and y-axis on the x-y plane.
The unit cell spacing is $\delta^\text{RIS,x}$ and $\delta^\text{RIS,y}$ in x and y directions, respectively.
The position of unit cell $(m,n)$ of RIS tile $k$ in the x-y plane of the local coordinate system is given by
\begin{align}
{\bold u}^\text{RIS}_{k,m,n} = (\delta^\text{RIS,x}\kappa_m^{M^\text{RIS}_k}, \delta^\text{RIS,y}\kappa_n^{N^\text{RIS}_k})^T.
\end{align}

The distance between the origin of the local coordinate systems (i.e., the center points) of the transmitter and RIS tile $k$ is denoted by $r^\text{Tx-RIS}_k$.
Similarly, $r^\text{RIS-Rx}_k$ is the distance between the origin of the local coordinate systems of RIS tile $k$ and the receiver.
We assume that the transmitter and receiver are located within the far-field region of each RIS tile.
That is, the size of an RIS tile is sufficiently small compared to the distances $r^\text{Tx-RIS}_k$ and $r^\text{RIS-Rx}_k$.
If this condition is not met, the RIS tile can be further divided into smaller RIS tiles.

The direction from the transmitter to RIS tile $k$ is represented by elevation $\theta^\text{Tx-RIS}_k$ and azimuth $\phi^\text{Tx-RIS}_k$ from the viewpoint of the local coordinate system of the transmitter.
On the other hand, the direction from RIS tile $k$ to the transmitter in the local coordinate system of RIS tile $k$ is represented by elevation $\theta^\text{RIS-Tx}_k$ and azimuth $\phi^\text{RIS-Tx}_k$.
In the same way, the directional relationship between the receiver and RIS tile $k$ is defined by $\theta^\text{Rx-RIS}_k$, $\phi^\text{Rx-RIS}_k$, $\theta^\text{RIS-Rx}_k$, and $\phi^\text{RIS-Rx}_k$.

In this paper, we prefer to use the u-v coordinates for representing the direction.
The u-v coordinate of elevation $\theta$ and azimuth $\phi$ is ${\bold s} = (\sin\theta\cos\phi, \sin\theta\sin\phi)^T$.
For example, the u-v coordinates of the directions from the transmitter to RIS tile $k$ is given by
\begin{align}
{\bold s}^\text{Tx-RIS}_k= (\sin\theta^\text{Tx-RIS}_k \cos\phi^\text{Tx-RIS}_k, \sin\theta^\text{Tx-RIS}_k \sin\phi^\text{Tx-RIS}_k)^T.
\end{align}
The u-v coordinates for other directions, denoted by ${\bold s}^\text{RIS-Tx}_k$, ${\bold s}^\text{Rx-RIS}_k$, and ${\bold s}^\text{RIS-Rx}_k$, can be defined in the same manner.

The spatial relationship between the transmitter and RIS tile $k$ is fully determined by three parameters $r^\text{Tx-RIS}_k$, ${\bold s}^\text{Tx-RIS}_k$, and ${\bold s}^\text{RIS-Tx}_k$, and that between the receiver and RIS tile $k$ is determined by $r^\text{Rx-RIS}_k$, ${\bold s}^\text{Rx-RIS}_k$, and ${\bold s}^\text{RIS-Rx}_k$.
We can describe the power transfer from the transmitter to the receiver through the RIS tiles based on these spatial relationship.
In this paper, we assume that the RIS tiles are positioned in such a way that no power is transferred between the RIS tiles.

The gains of the antenna elements in the transmitter and receiver towards RIS tile $k$ are denoted by $G_k^\text{Tx-RIS}$ and $G_k^\text{Rx-RIS}$, respectively.
Similarly, the gains of a unit cell of RIS tile $k$ towards the transmitter and receiver are denoted by $G_k^\text{RIS-Tx}$ and $G_k^\text{RIS-Rx}$, respectively.

\subsection{RIS Tile Control Model}\label{subsec:ris tile}

In this subsection, we model the unit cells in the RIS tiles, and explain the control mechanism of each RIS tile.
The incident free space wave is converted into a guided wave in the unit cell.
The incident guided wave is reflected by the controllable load with the normalized impedance of $z_L$, and the reflected guided wave is radiated to the air as the reflected free space wave.

The guided wave is reflected according to the following equation.
\begin{align}\label{eq:wre}
y = \Gamma x,
\end{align}
where $x$ and $y$ are the power waves of the incident and reflected guided wave of a unit cell and $\Gamma$ is the reflection coefficient such that
\begin{align}
\Gamma = \frac{z_L-1}{z_L+1}.
\end{align}
The unit cell and the controllable load are designed in such a way that the phase of the reflected wave is controlled by varying the reflection coefficient.
Then, the reflection coefficient is given by
\begin{align}
\Gamma = \exp(j\xi),
\end{align}
where $\xi$ is the phase shift.
The phase shift can be continuous, for example, if the varactor is used as a controllable load.
On the other hand, if a PIN diode is used as a controllable load, the phase shift is discrete.
Let $\Psi$ denote the number of the control states of the unit cell.
Then, the $\psi$th state of a unit cell corresponds to the phase shift of
\begin{align}
\xi = \frac{2\pi (\psi-1)}{\Psi},
\end{align}
for $\psi=1,\ldots,\Psi$.

We define $\Gamma_{k,m,n}$ as the reflection coefficient of unit cell $(m,n)$ of RIS tile $k$.
Since there are a large number of unit cells, it is difficult to find out the optimal reflection coefficient for each unit cell individually.
Instead, we introduce high-level control parameters for each RIS tile.
The direction of the reflected wave from RIS tile $k$ is controlled by elevation $\theta^\text{RIS}_k$ and azimuth $\phi^\text{RIS}_k$.
With these control parameters set, the incident wave coming from the normal direction of the RIS tile is reflected towards  the direction of elevation $\theta^\text{RIS}_k$ and azimuth $\phi^\text{RIS}_k$.
The u-v coordinate of $\theta^\text{RIS}_k$ and $\phi^\text{RIS}_k$ is given by
\begin{align}
{\bold c}_k= (\sin\theta^\text{RIS}_k \cos\phi^\text{RIS}_k, \sin\theta^\text{RIS}_k \sin\phi^\text{RIS}_k)^T.
\end{align}
We use ${\bold c}_k$ as the direction control parameter of RIS tile $k$.
In addition, the phase of the reflected wave is controlled by the phase control parameter $w_k$ for RIS tile $k$.

With the direction control parameter ${\bold c}$ and phase control parameter $w$, the reflection coefficient of each unit cell in the case of the continuous phase shift is set as follows.
\begin{align}\label{eq:riscontrol}
\begin{split}
\Gamma_{k,m,n}({\bold c},w) = \exp\big(j w\big)\exp\Big(-j\frac{2\pi}{\lambda}{\bold c}^T {\bold u}^\text{RIS}_{k,m,n} \Big).
\end{split}
\end{align}
If the phase shift is discrete, the reflection coefficient is quantized as follows.
\begin{align}\label{eq:quantized_riscontrol}
\begin{split}
\Gamma_{k,m,n}({\bold c},w) = \Omega_\Psi\bigg(\exp\big(j w\big)\exp\Big(-j\frac{2\pi}{\lambda}{\bold c}^T {\bold u}^\text{RIS}_{k,m,n} \Big)\bigg),
\end{split}
\end{align}
where $\Omega_\Psi$ is the phase quantization function.
For a complex number $z$, the phase quantization function is defined as
\begin{align}
\Omega_\Psi(z) = |z|\exp\bigg(j\frac{2\pi}{\Psi}\bigg\lfloor\frac{\Psi}{2\pi}\angle{z} + 0.5\bigg\rfloor\bigg),
\end{align}
where $|z|$ and $\angle{z}$ is the magnitude and phase of $z$.

\subsection{Antenna Array Transmitter and Receiver Model}

The transmitter is a phased antenna array, each antenna element of which has a phase shifter to control the phase of the transmitted wave.
The transmitter sends out equal power from all antenna elements.
Let $p^\text{Tx}$ denote the transmit power from one antenna element.
Similar to the RIS control, the transmitter is also controlled by high-level control parameters.
The direction of the transmitted wave is represented by elevation $\theta^\text{Tx}$ and azimuth $\phi^\text{Tx}$.
The direction control parameter of the transmitter is the u-v coordinate of $\theta^\text{Tx}$ and $\phi^\text{Tx}$ such that
\begin{align}
{\bold q}= (\sin\theta^\text{Tx} \cos\phi^\text{Tx}, \sin\theta^\text{Tx} \sin\phi^\text{Tx})^T.
\end{align}

Let $x_{i,j}$ denote the transmitted power wave from antenna element $(i,j)$ of the transmitter.
In addition, the transmitter excitation vector is defined as the vector of the transmitted power waves such that
\begin{align}
{\bold x} = (x_{i,j})_{i=1,\ldots,M^\text{Tx} \atop j=1,\ldots,N^\text{Tx}}.
\end{align}
When the direction control parameter ${\bold q}$ is given, the transmitted power wave from antenna element $(i,j)$ of the transmitter, which is denoted by $x_{i,j}({\bold q})$, is 
\begin{align}\label{eq:arraycontrol}
\begin{split}
x_{i,j}({\bold q}) = \sqrt{2p^\text{Tx}}\cdot \exp\Big(-j\frac{2\pi}{\lambda}{\bold q}^T {\bold u}^\text{Tx}_{i,j} \Big).
\end{split}
\end{align}
We define the directional transmitter excitation vector as
\begin{align}
{\bold x}({\bold q}) = (x_{i,j}({\bold q}))_{i=1,\ldots,M^\text{Tx} \atop j=1,\ldots,N^\text{Tx}}.
\end{align}

The receiver is an antenna array in which each antenna element receives the RF power.
We assume the DC combining technique for combining power from antenna elements.
In the DC combining technique, a rectifier attached to each antenna element performs RF-to-DC conversion, and then the DC power from each rectifier is summed together.
We define $\zeta$ as the RF-to-DC conversion function that maps the received RF power to the rectified DC power.
If $y_{a,b}$ denote the received power wave from antenna array $(a,b)$ of the receiver, the total received DC power $p^\text{Rx}_\text{DC}$ is given by
\begin{align}\label{eq:totpow}
p^\text{Rx}_\text{DC} = \sum_{a=1}^{M^\text{Rx}} \sum_{b=1}^{N^\text{Rx}} \zeta\bigg(\frac{|y_{a,b}|^2}{2}\bigg).
\end{align}
\section{Wireless Power Transfer Analysis}
\label{section:WPT}
\subsection{End-to-End Wireless Power Transfer Analysis}

In this section, we analyze the whole RIS-aided RF-WPT system based on the system model in the previous section.
We start with analyzing the channel gain between an antenna element of the transmitter or receiver to a unit cell of an RIS tile.
In general, the channel gain between two antennas are given by
\begin{align}\label{eq:h}
h = \frac{\lambda}{4\pi d}\sqrt{G^\text{A} G^\text{B}}\exp\bigg(-j\frac{2\pi}{\lambda}d\bigg),
\end{align}
where $d$ is the distance between two antennas and $G^\text{A}$ and $G^\text{B}$ are the antenna gains of two antennas.

Based on \eqref{eq:h}, we first derive the channel gain between antenna element $(i,j)$ of the transmitter to unit cell $(m,n)$ of RIS tile $k$.
Under the far-field assumption, the distance between the antenna element and unit cell is given by
\begin{align}\label{eq:dist}
d = r^\text{Tx-RIS}_k - ({\bold s}_k^\text{Tx-RIS})^T {\bold u}^\text{Tx}_{i,j} - ({\bold s}_k^\text{RIS-Tx})^T {\bold u}^\text{RIS}_{k,m,n}.
\end{align}
If we plug \eqref{eq:dist} into \eqref{eq:h} and replace the antenna gains with $G^\text{Tx-RIS}_k$ and $G^\text{RIS-Tx}_k$, we derive the channel gain from antenna element $(i,j)$ of the transmitter to unit cell $(m,n)$ of RIS tile $k$ as
\begin{align}\label{eq:htxris}
\begin{split}
&h_{(i,j),(k,m,n)}^\text{Tx-RIS}\\ 
&= \frac{\lambda}{4\pi r_k^\text{Tx-RIS}}\sqrt{G^\text{Tx-RIS}_k G^\text{RIS-Tx}_k}\exp\bigg(-j\frac{2\pi}{\lambda}r^\text{Tx-RIS}_k\bigg)\\
&\quad\times\exp\bigg(j\frac{2\pi}{\lambda}\Big(({\bold s}_k^\text{Tx-RIS})^T {\bold u}^\text{Tx}_{i,j} + ({\bold s}_k^\text{RIS-Tx})^T {\bold u}^\text{RIS}_{k,m,n}\Big)\bigg).\\
\end{split}
\end{align}
In a similar way, we derive the channel gain from unit cell $(m,n)$ of RIS tile $k$ to antenna element $(a,b)$ of the receiver as
\begin{align}\label{eq:hrisrx}
\begin{split}
&h_{(k,m,n),(a,b)}^\text{RIS-Rx}\\ 
&= \frac{\lambda}{4\pi r_k^\text{RIS-Rx}}\sqrt{G^\text{Rx-RIS}_k G^\text{RIS-Rx}_k}\exp\bigg(-j\frac{2\pi}{\lambda}r^\text{RIS-Rx}_k\bigg)\\
&\quad\times\exp\bigg(j\frac{2\pi}{\lambda}\Big(({\bold s}_k^\text{RIS-Rx})^T {\bold u}^\text{RIS}_{k,m,n}+({\bold s}_k^\text{Rx-RIS})^T {\bold u}^\text{Rx}_{a,b}\Big)\bigg).\\
\end{split}
\end{align}

From \eqref{eq:riscontrol}, \eqref{eq:htxris}, and \eqref{eq:hrisrx}, the channel gain from antenna element $(i,j)$ of the transmitter to antenna element $(a,b)$ of the receiver through unit cell $(m,n)$ of RIS tile $k$ is derived as
\begin{align}\label{eq:hijkmnab}
h_{(i,j),(k,m,n),(a,b)}^\text{Tx-RIS-Rx} = h_{(i,j),(k,m,n)}^\text{Tx-RIS} \Gamma_{k,m,n}({\bold c}_k,w_k) h_{(k,m,n),(a,b)}^\text{RIS-Rx}
\end{align}

Finally, the received power wave at antenna element $(a,b)$ of the receiver is given by
\begin{align}\label{eq:e2ewpt}
\begin{split}
&y_{a,b} = \\
&\sum_{i=1}^{M^\text{Tx}} \sum_{j=1}^{N^\text{Tx}}
\bigg(\sum_{k=1}^K \sum_{m=1}^{M^\text{RIS}_k} \sum_{n=1}^{N^\text{RIS}_k} 
h_{(i,j),(k,m,n),(a,b)}^\text{Tx-RIS-Rx} + h^\text{Tx-Rx}_{(i,j),(a,b)}\bigg) x_{i,j}({\bold q}),
\end{split}
\end{align}
where $h^\text{Tx-Rx}_{(i,j),(a,b)}$ is the direct channel gain from antenna element $(i,j)$ of the transmitter to antenna element $(a,b)$ of the receiver.

\subsection{Beam Steering Analysis}

Although the RF-WPT equation in \eqref{eq:e2ewpt} fully describes the end-to-end WPT, this equation is not intuitive since it treats the WPT as the summation of individual EM waves propagated via numerous antenna elements and unit cells.
Actually, the power is efficiently transferred when the beam is formed by the transmit antenna array and the RIS tiles.
By using the high level control methods of the RIS tile in \eqref{eq:riscontrol} and the transmit antenna array in \eqref{eq:arraycontrol}, we can reformulate \eqref{eq:e2ewpt} to describe the beam steering behavior from the transmitter and RIS tiles.

From \eqref{eq:riscontrol}, \eqref{eq:arraycontrol}, \eqref{eq:htxris}, and \eqref{eq:hrisrx}, we can calculate the received power wave relayed by RIS tile $k$ in \eqref{eq:e2ewpt} as
\begin{align}\label{eq:rxwavetilek}
\begin{split}
&\sum_{i=1}^{M^\text{Tx}} \sum_{j=1}^{N^\text{Tx}} \sum_{m=1}^{M^\text{RIS}_k} \sum_{n=1}^{N^\text{RIS}_k} h_{(i,j),(k,m,n),(a,b)}^\text{Tx-RIS-Rx} x_{i,j}({\bold q})\\
&= \bigg(\frac{\lambda}{4\pi}\bigg)^2\frac{\sqrt{G^\text{Tx-RIS}_k G^\text{RIS-Tx}_k G^\text{Rx-RIS}_k G^\text{RIS-Rx}_k (2p^\text{Tx})}}{r_k^\text{Tx-RIS}r_k^\text{RIS-Rx}}\\
&\quad\times\exp\bigg(-j\frac{2\pi}{\lambda}\big(r^\text{Tx-RIS}_k+r^\text{RIS-Rx}_k-({\bold s}_k^\text{Rx-RIS})^T {\bold u}^\text{Rx}_{a,b}\big)\bigg)\\
&\quad\times\exp\big(j w_k\big)\\
&\quad\times \sum_{m=1}^{M^\text{RIS}_k} \sum_{n=1}^{N^\text{RIS}_k} 
\exp\bigg(j\frac{2\pi}{\lambda}({\bold s}_k^\text{RIS-Tx}+{\bold s}_k^\text{RIS-Rx}-{\bold c}_k)^T {\bold u}^\text{RIS}_{k,m,n}\bigg)\\
&\quad\times\sum_{i=1}^{M^\text{Tx}} \sum_{j=1}^{N^\text{Tx}} \exp\bigg(j\frac{2\pi}{\lambda}({\bold s}_k^\text{Tx-RIS}-{\bold q})^T {\bold u}^\text{Tx}_{i,j}\bigg).
\end{split}
\end{align}

We can simplify \eqref{eq:rxwavetilek} as
\begin{align}\label{eq:rxwavetilek2}
\begin{split}
&R_{k,(a,b)} \exp\big(j w_k\big) U^\text{RIS}_k({\bold s}_k^\text{RIS-Tx}+{\bold s}_k^\text{RIS-Rx}-{\bold c}_k)\\
&\times U^\text{Tx}({\bold s}_k^\text{Tx-RIS}-{\bold q}).
\end{split}
\end{align}
In \eqref{eq:rxwavetilek2}, $R_{k,(a,b)}$ represents the magnitude and phase of the received power wave related to the distance throughout the transmitter, RIS tile $k$, and antenna element $(a,b)$ of the receiver, which is defined as 
\begin{align}
\begin{split}
&R_{k,(a,b)}\\
&= \bigg(\frac{\lambda}{4\pi}\bigg)^2\frac{\sqrt{G^\text{Tx-RIS}_k G^\text{RIS-Tx}_k G^\text{Rx-RIS}_k G^\text{RIS-Rx}_k (2p^\text{Tx})}}{r_k^\text{Tx-RIS}r_k^\text{RIS-Rx}}\\
&\quad\times\exp\bigg(-j\frac{2\pi}{\lambda}\big(r^\text{Tx-RIS}_k+r^\text{RIS-Rx}_k-({\bold s}_k^\text{Rx-RIS})^T {\bold u}^\text{Rx}_{a,b}\big)\bigg).
\end{split}
\end{align}

Furthermore, in \eqref{eq:rxwavetilek2}, $U^\text{RIS}_k({\bold v})$ denotes the beam steering function of RIS tile $k$ that maps a direction vector ${\bold v} = (v_x, v_y)^T$ in the u-v coordinate to the gain of the beam.
We define $U^\text{RIS}_k({\bold v})$ as
\begin{align}
\begin{split}
&U^\text{RIS}_k({\bold v})\\ 
&= M^\text{RIS}_k N^\text{RIS}_k \Xi_{M^\text{RIS}_k}\bigg(2\pi\frac{\delta^\text{RIS,x}}{\lambda}v_x\bigg)\Xi_{N^\text{RIS}_k}\bigg(2\pi\frac{\delta^\text{RIS,y}}{\lambda}v_y\bigg),
\end{split}
\end{align}
where $\Xi_M(x)$ is the periodic sinc function such that
\begin{align}\label{eq:ximx}
\begin{split}
    &\Xi_M(x)\\
    & = \frac{1}{M}\sum_{m=1}^{M} \exp\big(jx\kappa_{m}^{M}\big)\\
    & = 
    \begin{dcases}
        (-1)^{l(M-1)},\qquad\text{if }x = 2\pi l \text{ for } l=0,\pm 1,\pm 2, \ldots\\
        \frac{\sin(Mx/2)}{M\sin(x/2)},\qquad\text{otherwise}.
    \end{dcases}    
\end{split}
\end{align}
Since the periodic sinc function $\Xi_M(x)$ is maximized when $x$ equals to zero, $U^\text{RIS}_k({\bold v})$ has the maximum gain $M^\text{RIS}_k N^\text{RIS}_k$ if ${\bold v}$ is a zero vector.
This means that the gain of the beam from RIS tile $k$ to the receiver (i.e., $U^\text{RIS}_k({\bold s}_k^\text{RIS-Tx}+{\bold s}_k^\text{RIS-Rx}-{\bold c}_k)$ in \eqref{eq:rxwavetilek2}) is maximized when the direction control parameter of RIS tile $k$ is set as ${\bold c}_k = {\bold s}_k^\text{RIS-Tx}+{\bold s}_k^\text{RIS-Rx}$.

Similarly, $U^\text{Tx}({\bold v})$ denotes the beam steering function of the transmitter, defined as as
\begin{align}
\begin{split}
U^\text{Tx}({\bold v})
= M^\text{Tx} N^\text{Tx} \Xi_{M^\text{Tx}}\bigg(2\pi\frac{\delta^\text{Tx,x}}{\lambda}v_x\bigg)\Xi_{N^\text{Tx}}\bigg(2\pi\frac{\delta^\text{Tx,y}}{\lambda}v_y\bigg).
\end{split}
\end{align}
We can see that $U^\text{Tx}({\bold v})$ has the maximum gain $M^\text{Tx} N^\text{Tx}$ if ${\bold v}$ is a zero vector.
Therefore, the gain of the beam from the transmitter to RIS tile $k$ (i.e., $U^\text{Tx}({\bold s}_k^\text{Tx-RIS}-{\bold q})$ in \eqref{eq:rxwavetilek2}) is maximized when the direction control parameter of the transmitter is ${\bold q}={\bold s}_k^\text{Tx-RIS}$.

We define the set of the phase and direction control parameters for all RIS tiles as ${\boldsymbol \omega} = (\omega_1,\ldots,\omega_K)$ and ${\bold C} = ({\bold c}_1,\ldots,{\bold c}_K)$.
When the direction control is used for the transmitter and all RIS tiles, from \eqref{eq:rxwavetilek2}, the received wave at antenna element $(a,b)$ in \eqref{eq:e2ewpt} is rewritten as
\begin{align}
\begin{split}\label{eq:yab2}
&y_{a,b}({\boldsymbol \omega}, {\bold C}, {\bold q})\\ 
&= \sum_{k=1}^K R_{k,(a,b)} \exp\big(j w_k\big) U^\text{RIS}_k({\bold s}_k^\text{RIS-Tx}+{\bold s}_k^\text{RIS-Rx}-{\bold c}_k)\\
&\qquad\quad\times U^\text{Tx}({\bold s}_k^\text{Tx-RIS}-{\bold q})\\
&\quad+ \Phi_{a,b}({\bold q}),
\end{split}
\end{align}
where $\Phi_{a,b}({\bold q})$ is the power directly transferred to antenna element $(a,b)$ of the receiver without going through any unit cell of any RIS tile.
We define $\Phi_{a,b}({\bold q})$ as
\begin{align}
\Phi_{a,b}({\bold q}) = \sum_{i=1}^{M^\text{Tx}} \sum_{j=1}^{N^\text{Tx}} h^\text{Tx-Rx}_{(i,j),(a,b)} x_{i,j}({\bold q}).
\end{align}
We can calculate the total received DC power at the receiver from \eqref{eq:totpow} and \eqref{eq:yab2}.

The equation \eqref{eq:yab2} can be intuitively interpreted as follows.
The transmitter directs the beam to some selected RIS tiles by controlling the direction control parameter ${\bold q}$.
For these selected RIS tiles, and the gains of the beams from the transmitter to RIS tiles (i.e., $U^\text{Tx}({\bold s}_k^\text{Tx-RIS}-{\bold q})$) become high.
RIS tile $k$ reflects the beam from the transmitter towards some direction according to the direction control parameter ${\bold c}_k$.
RIS tile $k$ should set ${\bold c}_k$ as ${\bold c}_k = {\bold s}_k^\text{RIS-Tx}+{\bold s}_k^\text{RIS-Rx}$ to direct the reflected beam to the receiver.
The receiver receives the beams reflected from multiple RIS tiles as well as the wave directly transferred from the transmitter.
The phases of the beams from RIS tiles vary according to $R_{k,(a,b)}$, and they do not match the phase of the direct wave from the transmitter.
Therefore, each RIS tile should control the phase control parameter $w_k$ to align the phases of the beams at the receiver so that all the waves are optimally combined at the receiver.

\section{Power-Based Beam Scanning Algorithm}\label{section:powerscanning}

\subsection{Beam Scanning of Transmitter and RIS Tiles}
\label{subsec:scanningbeam}

In this subsection, we design the power-based beam scanning algorithm that controls the transmitter and RIS tiles to maximize the total receive power at the receiver.
In this algorithm, the transmitter and RIS tiles perform the beam scanning, and the receiver measures the receive power of each beam.
Then, the algorithm decides the best beam for the transmitter and RIS tiles based on the receive power measurements in an iterative manner.

The set of scanning beams for the transmitter and RIS tiles are generated by a uniform grid in the u-v coordinate system.
Each scanning beam is determined by a direction control parameter, ${\bold q}$ for the transmitter and ${\bold c}_k$ for RIS tile $k$.
Let ${\bold q}^{(l)}$ and ${\bold c}^{(l)}_k$ as the direction control parameter for the $l$th scanning beam of the transmitter and RIS tile $k$, respectively.
Then, we define ${\bold q}^{(l)}$ as
\begin{align}
{\bold q}^{(l)} = \bigg(\frac{\Delta^\text{Tx,u}}{\Upsilon^\text{Tx,u}-1}\kappa_{l^u}^{\Upsilon^\text{Tx,u}},
\frac{\Delta^\text{Tx,v}}{\Upsilon^\text{Tx,v}-1}\kappa_{l^v}^{\Upsilon^\text{Tx,v}}\bigg)^T,
\label{eq:Txscanbeam}
\end{align}
where $\Delta^\text{Tx,u}$ and $\Delta^\text{Tx,v}$ are, respectively, the scan widths along the u and v axes, $\Upsilon^\text{Tx,u}$ and $\Upsilon^\text{Tx,v}$ are, respectively, the numbers of scanning beams along the u and v axes, and $l^u = ((l-1) \pmod{\Upsilon^\text{Tx,u})}+1$ and $l^v = \lfloor(l-1)/\Upsilon^\text{Tx,u}\rfloor + 1$ are respectively the indices of the scanning beam along the u and v axes.
The number of scanning beams of the transmitter is $L^\text{Tx} = \Upsilon^\text{Tx,u}\Upsilon^\text{Tx,v}$.
Similarly, we define ${\bold c}^{(l)}_k$ as
\begin{align}
{\bold c}^{(l)}_k = \bigg(\frac{\Delta^\text{RIS,u}_k}{\Upsilon^\text{RIS,u}_k-1}\kappa_{l^u}^{\Upsilon^\text{RIS,u}_k},
\frac{\Delta^\text{RIS,v}_k}{\Upsilon^\text{RIS,v}_k-1}\kappa_{l^v}^{\Upsilon^\text{RIS,v}_k}\bigg)^T,
\label{eq:RISscanbeam}
\end{align}
where $\Delta^\text{RIS,u}_k$ and $\Delta^\text{RIS,v}_k$ are, respectively, the scan widths along the u and v axes for RIS tile $k$, $\Upsilon^\text{RIS,u}_k$ and $\Upsilon^\text{RIS,v}_k$ are, respectively, the number of scanning beams along the u and v axes for RIS tile $k$, and $l^u = ((l-1) \pmod{\Upsilon^\text{RIS,u}_k)}+1$ and $l^v = \lfloor(l-1)/\Upsilon^\text{RIS,v}_k\rfloor + 1$ are, respectively, the indices of the scanning beam along the u and v axes.
The number of scanning beams of RIS tile $k$ is $L_k^\text{RIS} = \Upsilon_k^\text{RIS,u}\Upsilon_k^\text{RIS,v}$.

The receiver designates one antenna element at the center of the antenna array as a sensor antenna, and measures the receive power of the sensor antenna.
Let antenna element $(\bar{a},\bar{b})$ be the sensor antenna.
Then, when the direction control is used for the transmitter and all RIS tiles, the receive power measurement is given by
\begin{align}
P = \frac{|y_{\bar{a},\bar{b}}|^2}{2} = \frac{|\sum_{k=1}^K\exp(j\omega_k)\alpha_k({\bold c}_k)\beta_k({\bold q}) + \gamma({\bold q})|^2}{2},
\end{align}
where
\begin{align}
\alpha_k({\bold c}) &= R_{k,(\bar{a},\bar{b})} U^\text{RIS}_k({\bold s}_k^\text{RIS-Tx}+{\bold s}_k^\text{RIS-Rx}-{\bold c}),\\
\beta_k({\bold q}) &= U^\text{Tx}({\bold s}_k^\text{Tx-RIS}-{\bold q}),\\
\gamma({\bold q}) &= \Phi_{\bar{a},\bar{b}}({\bold q}).
\end{align}

To find out the best beam towards RIS tiles, the transmitter can generate a wide beam by turning on only one antenna element $(\bar{i}, \bar{j})$ while turning off all other antenna elements.
In this case, the following wide-beam transmitter excitation vector is used by the transmitter.
\begin{align}\label{eq:widebeamparam}
\widehat{\bold x} = (\widehat{x}_{i,j})_{i=1,\ldots,M^\text{Tx} \atop j=1,\ldots,N^\text{Tx}},
\end{align}
where $\widehat{x}_{\bar{i}, \bar{j}} = \sqrt{2p^\text{Tx}}$ for antenna element $(\bar{i}, \bar{j})$ and $\widehat{x}_{i,j} = 0$ for all $(i,j)$ except for $(\bar{i}, \bar{j})$.
Then, the receive power measurement is simplified to
\begin{align}
P = \frac{|\sum_{k=1}^K\exp(j\omega_k)\alpha_k({\bold c}_k)\bar{\beta}_k + \bar{\gamma}|^2}{2},
\label{eq:oneTx}
\end{align}
where
\begin{align}
\bar{\beta}_k &= \exp\bigg(j\frac{2\pi}{\lambda}({\bold s}_k^\text{Tx-RIS})^T {\bold u}^\text{Tx}_{\bar{i},\bar{j}}\bigg),\\
\bar{\gamma} &= h^\text{Tx-Rx}_{(\bar{i},\bar{j}),(\bar{a},\bar{b})} \sqrt{2p^\text{Tx}}.
\end{align}

\subsection{RIS Tile and Transmitter Scanning Method} \label{subsec:tilescan}
In this subsection, we propose the RIS tile scanning method that finds the best phase and direction control parameters for a specific RIS tile while the transmitter and other RIS tiles are fixed.
Let us suppose that the scanning for RIS tile $\bar{k}$ is performed.
To focus on RIS tile $\bar{k}$, the receive power measurement is rewritten as
\begin{align}\label{eq:powmea}
P(\omega, {\bold c}) = \frac{|X + \exp(j\omega)Y({\bold c})|^2}{2},
\end{align}
where $\omega$ and ${\bold c}$ are, respectively, the phase and direction control parameters for RIS tile $\bar{k}$, and
\begin{align}
X &= \sum_{k=1,\ldots,K \atop k\ne \bar{k}}\exp(j\omega_k)\alpha_k({\bold c}_k)\beta_k({\bold q}) + \gamma({\bold q})\\
Y({\bold c}) &= \alpha_{\bar{k}}({\bold c})\beta_{\bar{k}}({\bold q}),
\end{align}
if the direction control is used for the transmitter, and
\begin{align}
X &= \sum_{k=1,\ldots,K \atop k\ne \bar{k}}\exp(j\omega_k)\alpha_k({\bold c}_k)\bar{\beta}_k + \bar{\gamma}\\
Y({\bold c}) &= \alpha_{\bar{k}}({\bold c})\bar\beta_{\bar{k}},
\end{align}
if only one antenna element of the transmitter is turned on.

\begin{figure}[!htb]
    \centering
    \includegraphics[trim = 5cm 2.5cm 1cm 2cm, clip = true,width = 0.45\textwidth]{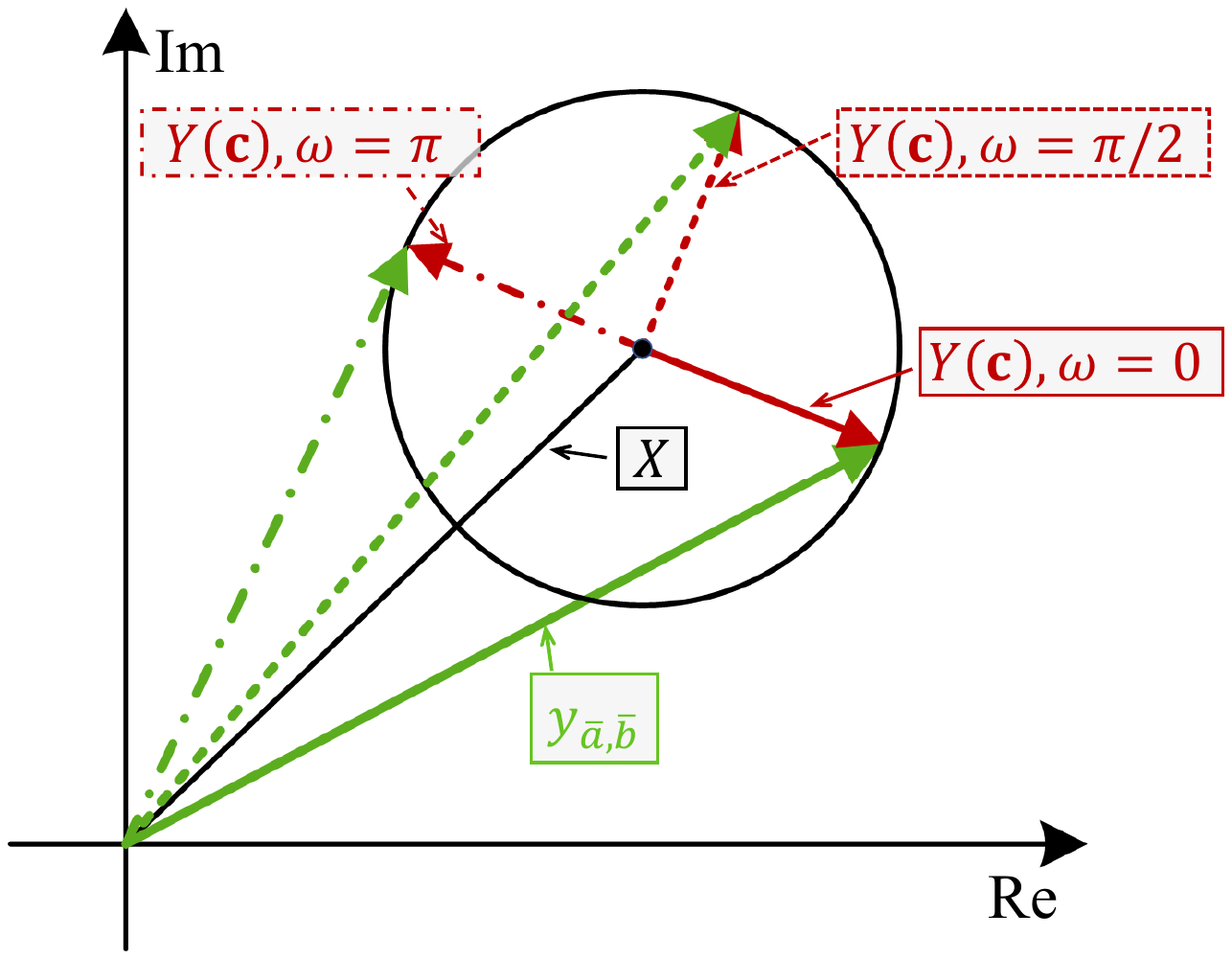}
    \caption{Receive signal with different phase control parameters.}
    \label{fig:RIStilescan}
\end{figure}

Fig.~\ref{fig:RIStilescan} shows the receive signal $y_{\overline{a},\overline{b}}$ for three different phase control parameters for RIS tile $\bar{k}$. 
It is clear that the receive signal relayed by RIS tile $\bar{k}$ (i.e., $\exp(j\omega)Y({\bold c})$) is rotated with respect to the phase control parameter $\omega$. 
In contrast, the signal $X$ relayed from other RIS tiles and directly received from the transmitter are fixed.

For the given direction control parameter ${\bold c}$, the optimal phase control parameter aligns the phases of $X$ and $Y$.
The optimal phase control parameter is given by
\begin{align}\label{eq:omegaopt}
\omega^\text{opt}({\bold c}) = \angle{(XY({\bold c})^*)},
\end{align}
and the receive power for the optimal phase control parameter is
\begin{align}\label{eq:popt}
\begin{split}
P(\omega^\text{opt}({\bold c}), {\bold c}) &= \frac{(|X| + |Y({\bold c})|)^2}{2}\\ 
&= \frac{|X|^2 + |Y({\bold c})|^2}{2} + |XY({\bold c})^*|.
\end{split}
\end{align}

To compute the optimal phase control parameter $\omega^\text{opt}({\bold c})$, we need to measure the receive power with three different phase control parameters, $\omega = 0,~\pi,\text{ and }\pi/2$ as in Fig.~\ref{fig:RIStilescan}.
From \eqref{eq:powmea}, we have the receive power measurements for these phase control parameters as
\begin{align}
&P(0, {\bold c}) = (|X|^2 + |Y({\bold c})|^2 + X^*Y({\bold c}) + XY({\bold c})^*)/2,\\
&P(\pi, {\bold c}) = (|X|^2 + |Y({\bold c})|^2 - X^*Y({\bold c}) - XY({\bold c})^*)/2,\\
&P(\pi/2, {\bold c}) = (|X|^2 + |Y({\bold c})|^2 + jX^*Y({\bold c}) - jXY({\bold c})^*)/2.
\end{align}
By solving these linear equations, we have
\begin{align}
&|X|^2 + |Y({\bold c})|^2 = P(0, {\bold c}) + P(\pi, {\bold c}),\label{eq:x2y2}\\
&XY({\bold c})^* = \frac{1-j}{2}P(0, {\bold c}) -\frac{1+j}{2} P(\pi, {\bold c}) + j P(\pi/2, {\bold c}).\label{eq:xyc}
\end{align}

From \eqref{eq:omegaopt}, \eqref{eq:popt}, \eqref{eq:x2y2}, and \eqref{eq:xyc}, the optimal phase control parameter and the corresponding receive power are calculated based on the receive power measurements as
\begin{align}
\omega^\text{opt}({\bold c}) = \angle{\bigg(\frac{1-j}{2}P(0, {\bold c}) -\frac{1+j}{2} P(\pi, {\bold c}) + j P(\pi/2, {\bold c})\bigg)},
\label{eq:omegaopt2}
\end{align}
and
\begin{align}
\begin{split}
&P(\omega^\text{opt}({\bold c}), {\bold c}) = \frac{P(0, {\bold c}) + P(\pi, {\bold c})}{2}\\ 
&\qquad + \bigg|\frac{1-j}{2}P(0, {\bold c}) -\frac{1+j}{2} P(\pi, {\bold c}) + j P(\pi/2, {\bold c})\bigg|,
\end{split}
\label{eq:calpow}
\end{align}
respectively.

Now, we develop an algorithm that finds out the optimal direction control parameter and the corresponding phase control parameter that maximize the receive power $P(\omega^\text{opt}({\bold c}), {\bold c})$ in \eqref{eq:calpow}.
In Algorithm \ref{RISTileScan}, we propose the RIS tile scanning algorithm for a target RIS tile. 
Algorithm \ref{RISTileScan} scans the control parameters of the target RIS tile while keeping those of other RIS tiles and the transmitter unchanged.
The index of the target RIS tile is denoted by $\bar{k}$.
We denote by ${\bold C}_{-\bar{k}}$ and ${\boldsymbol \omega}_{-\bar{k}}$ the vectors containing the direction and phase control parameters of the RIS tiles other than the target RIS tile, which is given in Algorithm \ref{RISTileScan} as an input.
In addition, $\bold x$ is the transmitter excitation vector, which is also given in the algorithm as an input. 
The algorithm starts by scanning the direction control parameters of the target RIS tile with $L^{\text{RIS}}_{\bar{k}}$ scanning beams (line 3) while keeping the parameters of the other RIS tiles fixed to ${\bold C}_{-\bar{k}}$ and ${\boldsymbol \omega}_{-\bar{k}}$ (line 1) and the parameters of the transmitter fixed to $\bold x$ (line 2). 
For each scanning beam, the algorithm changes the phase control parameter among three values (i.e., $0$, $\pi$, and $\pi/2$) and measures the corresponding receive power at line 4 of the algorithm. 
Then, the optimal phase control parameter $\omega^{(l)}$ and the corresponding receive power $P^{(l)}$ for the $l$th direction control parameter are computed at lines 5--7 according to \eqref{eq:omegaopt2} and \eqref{eq:calpow}.
The index of the scanning beam with the highest receive power is figured out at line 9. 
Then, the optimal direction and phase control parameters resulting in the highest receive power are returned at lines 10--12. 

\begin{algorithm}[!ht]
\SetAlgoLined
\KwIn{
\\Index of target RIS tile $\bar{k}$\\
Direction control parameters of other RIS tiles ${\bold C}_{-\bar{k}}$\\
Phase control parameters of other RIS tiles ${\boldsymbol \omega}_{-\bar{k}}$\\
Transmitter excitation vector ${\bold x}$
}
\KwOut{
\\Optimal direction control parameter ${\bold c}^\text{opt}$\\ 
Optimal phase control parameter $\omega^\text{opt}$
}
Set the direction and phase control parameters of all RIS tiles except for the target RIS tile according to ${\bold C}_{-\bar{k}}$ and ${\boldsymbol \omega}_{-\bar{k}}$\\
Set the transmitter excitation vector ${\bold x}$ to the transmitter\\
\BlankLine
\For{$l \leftarrow 1$ \KwTo $L^\text{\emph{RIS}}_{\bar{k}}$}{
Measure $P(0, {\bold c}_{\bar{k}}^{(l)})$, $P(\pi, {\bold c}_{\bar{k}}^{(l)})$, and $P(\frac{\pi}{2}, {\bold c}_{\bar{k}}^{(l)})$\\
${\boldsymbol \vartheta} \leftarrow \frac{1-j}{2}P(0, {\bold c}_{\bar{k}}^{(l)}) - \frac{1+j}{2}P(\pi, {\bold c}_{\bar{k}}^{(l)})+jP(\frac{\pi}{2}, {\bold c}_{\bar{k}}^{(l)})$\\
$\omega^{(l)} \leftarrow \angle{\boldsymbol \vartheta}$\\
$P^{(l)} \leftarrow \frac{P(0, {\bold c}_{\bar{k}}^{(l)})+P(\pi, {\bold c}_{\bar{k}}^{(l)})}{2}+|{\boldsymbol \vartheta}|$
}
$l^\text{opt} \leftarrow \arg\max_{l=1,\ldots,L^{\text{RIS}}_{\bar{k}}}{P^{(l)}}$\\
${\bold c}^\text{opt} \leftarrow {\bold c}^{(l^\text{opt})}_{\bar{k}}$\\
$\omega^\text{opt} \leftarrow \omega^{(l^\text{opt})}$\\
\Return ${\bold c}^\text{opt}$, $\omega^\text{opt}$
\caption{RIS Tile Scanning Algorithm} \label{RISTileScan}
\end{algorithm}

The transmitter should be optimized as well as the RIS tiles.
In Algorithm \ref{TxScan}, we present a simple transmitter scanning algorithm to find the optimal direction control parameter of the transmitter. 
The vectors of direction and phase control parameters of all RIS tiles (i.e., ${\bold C}$ and $\boldsymbol \omega$) are given in the algorithm as an input. 
The algorithm scans the beams of the transmitter with $L^\text{Tx}$ scanning beams (lines 2-5) while the whole RIS is loaded with the RIS direction and phase control parameter vectors ${\bold C}$ and $\boldsymbol \omega$ (line 1).
The optimal direction control parameter with the maximum receive power is returned at lines 6-8.

\begin{algorithm}[!ht]
\SetAlgoLined
\KwIn{
\\RIS direction control parameter vector ${\bold C}$\\ 
RIS phase control parameter vector ${\boldsymbol \omega}$
}
\KwOut{
\\ Optimal transmitter direction control parameter ${\bold q}^\text{opt}$
}
\BlankLine
Set the direction and phase control parameters of all RIS tiles to ${\bold C}$ and ${\boldsymbol \omega}$\\
\For{$l \leftarrow 1$ \KwTo $L^\text{Tx}$}{
Set the transmitter excitation vector to ${\bold x}({\bold q}^{(l)})$\\
Measure the receive power $P^{(l)}$\\
}
$l^\text{opt}\leftarrow \arg\max_{l=1,\ldots, L^{\text{Tx}}}P^{(l)}$\\
${\bold q}^\text{opt} \leftarrow {\bold q}^{(l^\text{opt})}$\\

\Return ${\bold q}^\text{opt}$
\caption{Transmitter Scanning Algorithm} \label{TxScan}
\end{algorithm}

\subsection{Multi-Tile RIS Beam Scanning Algorithm} \label{subsec:MTBSalgorithm}

In this subsection, we introduce the multi-tile RIS beam scanning (MTBS) algorithm to get the optimal control parameters for all RIS tiles and the transmitter with only receive power information. 
To reduce the scanning time, we propose a smart beam scanning algorithm that alternately scans and optimizes the RIS tiles and transmitter by Algorithms \ref{RISTileScan} and \ref{TxScan}.
The overall operation of the MTBS algorithm is summarized as follows.
In the first iteration, only one antenna element of the transmitter is turned on for generating a wide beam from the transmitter. 
Then, all RIS tiles are scanned and optimized one by one with the RIS tile scanning algorithm. 
After setting all RIS tiles with current optimal control parameters, the beams for the transmitter are scanned to figure out the best direction control parameter of the transmitter. 
From the next iteration, the transmitter activates all antenna elements and uses the best direction control parameter.
Then, we repeatedly perform the scanning and optimizing for all RIS tiles and the transmitter to enhance the receive power over iterations. 

\begin{algorithm}[!ht]
\SetAlgoLined
\KwIn{
\\ Number of scanning iterations $\eta$
}
\KwOut{
\\ Optimal RIS direction control parameter vector ${\bold C}^{\text{opt}}$\\ 
Optimal RIS phase control parameter vector ${\boldsymbol \omega}^{\text{opt}}$\\
Optimal transmitter direction control parameter ${\bold q}^{\text{opt}}$\\
}
${\bold x} \leftarrow \widehat{\bold x}$\\
\For{$\tau\leftarrow 1$ \KwTo $\eta$}{
\For{$k \leftarrow 1$ \KwTo $K$}{
$({\bold c}_k,{\omega}_k) \leftarrow \text{RIS tile scanning algorithm}(k,{\bold C}_{-k}, {\boldsymbol \omega}_{-k},{\bold x})$\\
}
${\bold q} \leftarrow$ Transmitter scanning algorithm$({\bold C}, \boldsymbol \omega)$\\
${\bold x} \leftarrow {\bold x}({\bold q})$\\
}
${\bold C}^{\text{opt}} \leftarrow {\bold C}$\\
${\boldsymbol \omega}^{\text{opt}} \leftarrow {\boldsymbol \omega}$\\
${\bold q}^{\text{opt}} \leftarrow {\bold q}$\\
\Return ${\bold C}^{\text{opt}}$, ${\boldsymbol \omega}^{\text{opt}}$, ${\bold q}^{\text{opt}}$
\caption{Multi-tile RIS Beam Scanning (MTBS) Algorithm} \label{Scanning}
\end{algorithm}

The details of the MTBS algorithm is presented in Algorithm \ref{Scanning}. 
This algorithm maintains and updates the RIS direction and phase control parameter vectors (i.e. ${\bold C}$ and ${\boldsymbol \omega}$) and the transmitter direction control parameter and excitation vector (i.e., ${\bold q}$ and ${\bold x}$).
This algorithm starts by activating one antenna element of the transmitter with the wide-beam transmitter excitation vector $\widehat{{\bold x}}$ in \eqref{eq:widebeamparam} at line 1. 
Then, each RIS tile is optimized by the RIS tile scanning algorithm in Algorithms \ref{RISTileScan} at lines 3-5. 
Subsequently, the direction control parameter of the transmitter is updated by the transmitter scanning algorithm in Algorithm \ref{TxScan} at line 7. 
The corresponding transmitter excitation vector (i.e., ${\bold x}({\bold q})$) is updated to ${\bold x}$ at line 6. 
This process is repeated $\eta$ times, and the optimal control parameters are obtained.

\section{RIS-aided WPT System Implementation}\label{sec:implementation}  

\subsection{RIS Design and Fabrication}
\begin{figure}[!htb]
    \centering
    \includegraphics[trim = 6.5cm 5.5cm 8.25cm 4.5cm, clip =true, width = 0.48\textwidth]{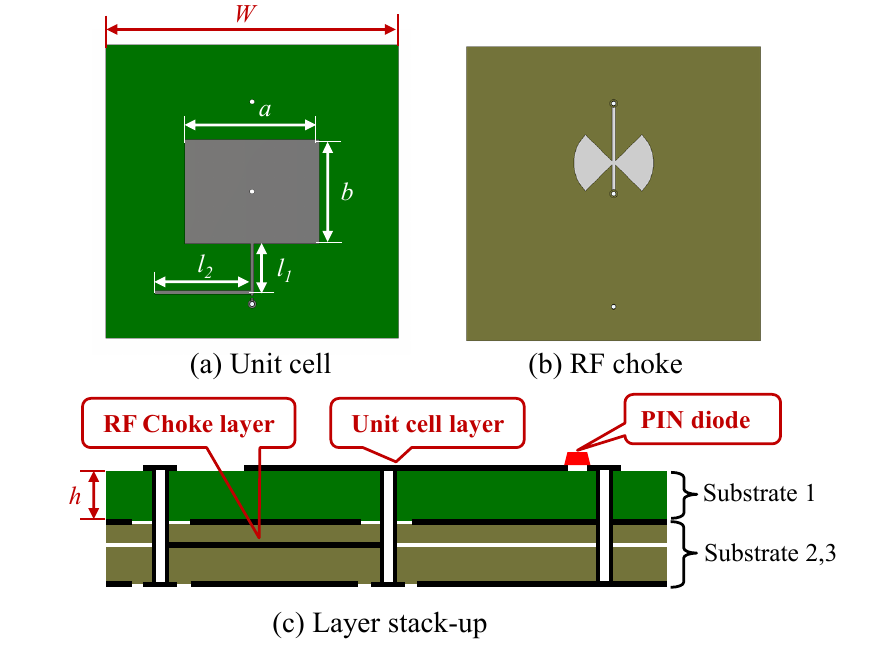}
    \caption{Unit cell configuration.}
    \label{fig:RISunitcell}
\end{figure}

In this section, we present the design and fabrication of the RIS. We first design a 1-bit guided wave unit cell which operates at 5.8 GHz. The configuration of the unit cell is given in Fig.~\ref{fig:RISunitcell}. The unit cell is designed on Rogers RO4350B substrate (i.e., substrate 1 in Fig.~\ref{fig:RISunitcell}(c)) with the relative permittivity of 3.66 and thickness of 1.524 mm. Furthermore, an RF choke placed on the FR4-epoxy substrate (i.e., substrate 2 in Fig.~\ref{fig:RISunitcell}(c)) is incorporated into the unit cell.

\begin{figure}[!htb]
    \centering
    \includegraphics[trim = 2cm 8cm 12cm 3cm, clip =true, width = 0.4\textwidth]{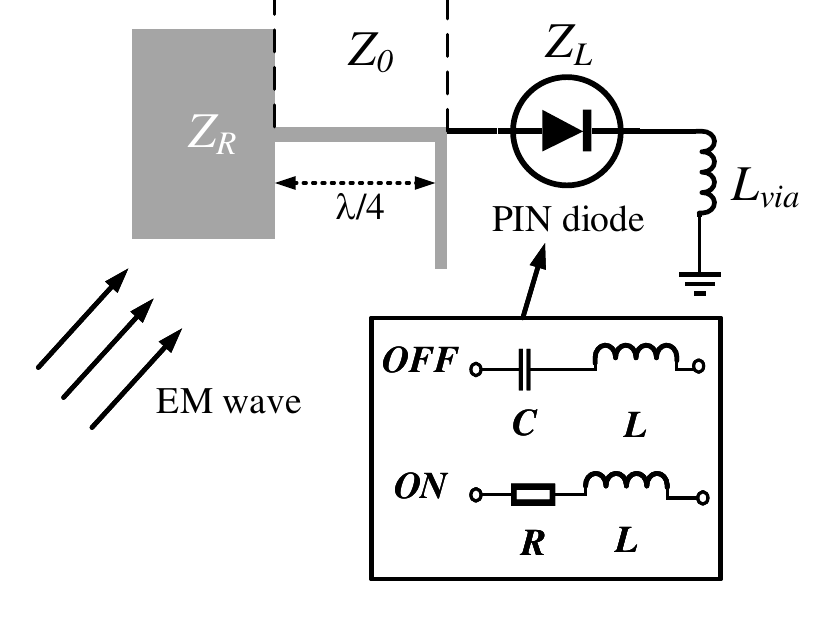}
    \caption{Equivalent circuit of the 1-bit guided wave unit cell.}
    \label{fig:RISequivalent}
\end{figure}

The principle of designing the unit cell is given in Fig.~\ref{fig:RISequivalent}. Firstly, a rectangular patch is designed to resonate at 5.8 GHz. 
The EM wave at the resonant frequency is absorbed by the path with the radiation impedance of $Z_R$. 
Then, this wave is guided to a quarter wavelength transmission line in the form of a guided wave. 
The quarter wavelength transmission line with characteristic impedance $Z_0$ in combination with a stub is carefully designed to transform the radiation impedance into an appropriate value (e.g., transformed impedance $Z_t$). 
The guided wave is reflected at the controllable load (i.e., a PIN diode) with the impedance $Z_L$. 
Afterward, this reflected guided wave propagates back to the patch and is re-radiated to the air. 
The reflection coefficient at the PIN diode can be expressed as
\begin{equation}
    \Gamma = \frac{Z_L-Z_t}{Z_L+Z_t} = \frac{z_L-1}{z_L+1}=|\Gamma|\exp{j\xi},
    \label{eq14}
\end{equation}
where 
\begin{align}
    Z_L = \begin{cases} 
        R + j\omega L & \quad \text{ON state}, \\
        \frac{1}{j\omega C}+j\omega L &\quad \text{OFF state},
    \end{cases}
\end{align}
$Z_t = Z_0^2/Z_R$ is the transformed impedance, $z_L = Z_L/Z_t$ is the normalized impedance, $\xi$ is the phase shift. Ideally, the magnitude of the reflection coefficient (i.e., $|\Gamma|$) equals to 1. 
To get a $180^\circ$ phase difference, one should figure out the values of $Z_t$ and $Z_0$ which satisfy
\begin{equation}
    \Delta\xi = \xi_{ON}-\xi_{OFF} = 180^\circ.
    \label{eq15}
\end{equation}

In this design, we use a PIN diode (MACOM MADP-000907-14020) to control the reflection phase of the guided wave. 
The PIN diode can be modeled as a series RL circuit (R = 5.2 $\Omega$, L = 30 pH) in ON state, and as a series LC circuit (L = 30 pH, C = 30 fF) in OFF state. 
To ensure the PIN diode operation, a biasing circuit with an RF choke is incorporated (see Fig.~\ref{fig:RISunitcell}(b)).
The RF choke is designed with a butterfly stub to prevent high-frequency signals from traveling to the DC source. 
To minimize the biasing circuit loss, we design the biasing point at the center of the patch, which is a zero electric field point. 
Consequently, good isolation and no additional loss from the DC biasing circuit to the main patch is achieved. 
The geometry parameters of the unit cell are listed in Table~\ref{table1}.

\begin{table}[!h]
\caption{Geometry parameters of the unit cell (unit: mm).}
\label{table1}
\centering
\begin{tabular}{cccc}
\hline\hline
\textbf{Parameter} & \textbf{Value} & \textbf{Parameter} & \textbf{Value} \\ \hline
\textit{W}                  & 36.2           & $l_1$                 & 7.5            \\ \hline
\textit{a}                  & 16.61          & $l_2$                 & 12             \\ \hline
\textit{b}                  & 12.88          & \textit{h}                  & 1.524          \\ \hline\hline
\end{tabular}
\end{table}

\begin{figure}[!htb]
    \centering
    \includegraphics[trim = 2cm 1cm 0.5cm 2cm,clip = true, width = 0.48\textwidth]{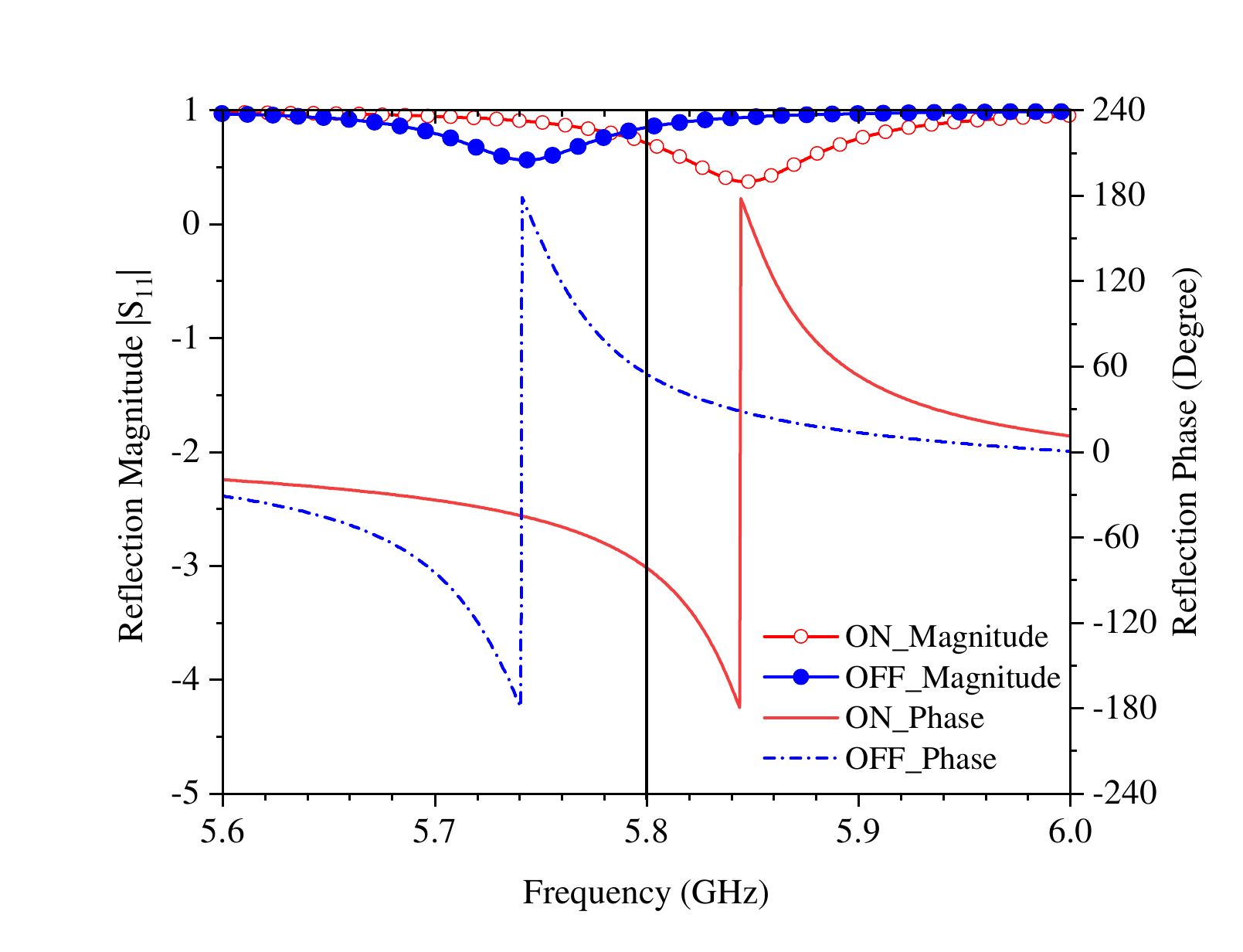}
    \caption{Simulation result of the unit cell in ON/OFF states.}
    \label{fig:RISsimresult}
\end{figure}

The unit cell performance is validated by simulating in an EM simulator with periodic boundary and Floquet port excitation. 
Fig.~\ref{fig:RISsimresult} indicates the magnitude and phase of the simulated reflection coefficient. 
It is evident that a nearly $180^\circ$ phase shift between ON and OFF state is achieved within 100 MHz bandwidth around the center frequency. 
The reflection amplitude is almost uniform between the two states. 
A small amount of loss can be observed due to the lossy substrate and the heat generated by the resistance of the PIN diode in ON state.

\begin{figure}[!htb]
    \centering
    \includegraphics[trim = 6cm 3cm 6cm 2cm, clip = true,width =0.42\textwidth]{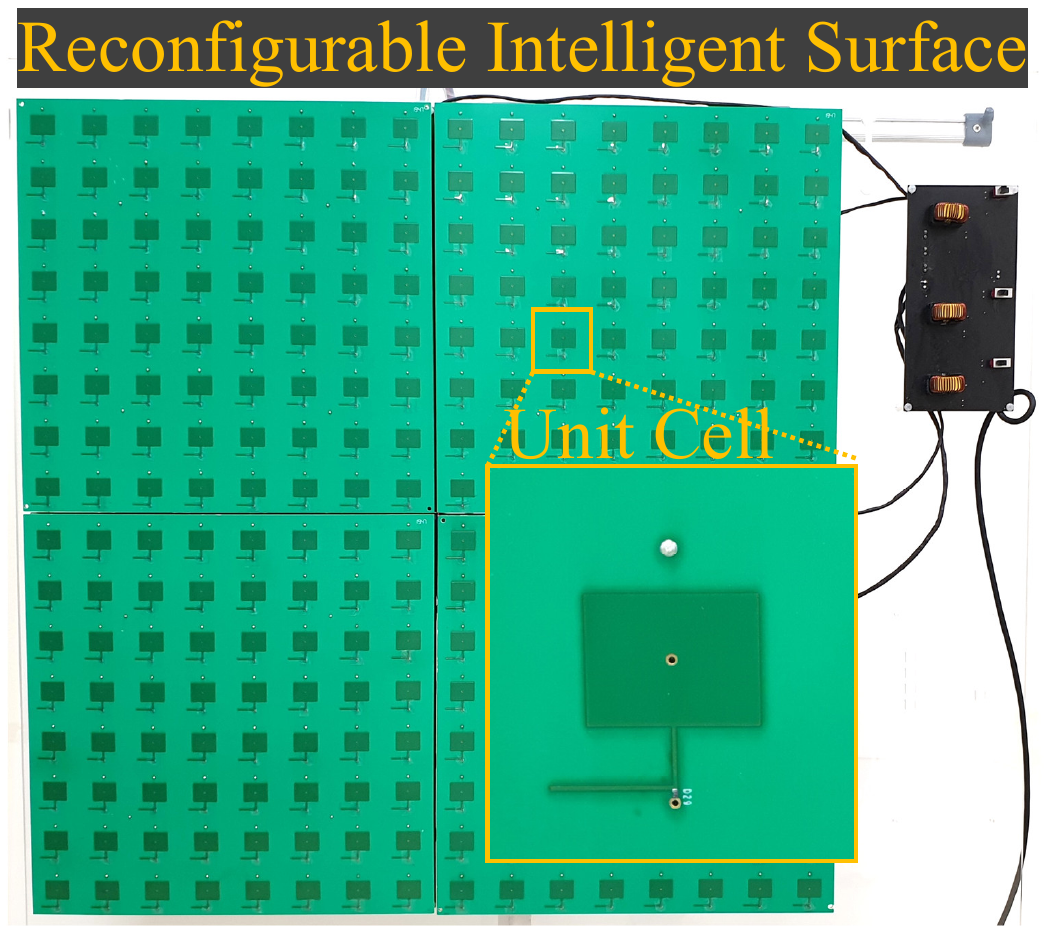}
    \caption{$16\times16$ RIS prototype.}
    \label{fig:RIS_board}
\end{figure}

\begin{figure}[!htb]
    \centering
    \includegraphics[trim = 3.75cm 0.5cm 5.75cm 0.5cm, clip = true,width =0.48\textwidth]{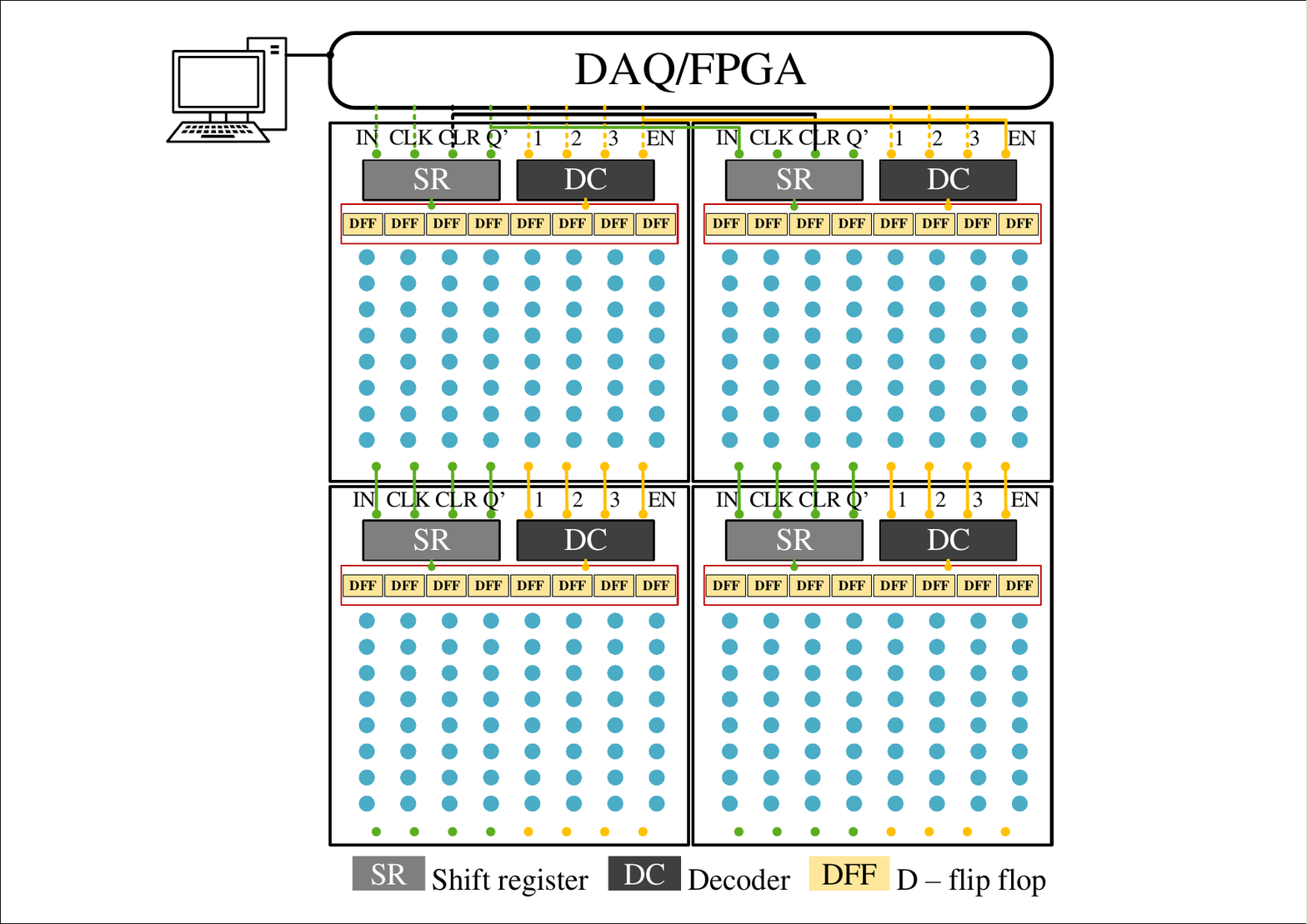}
    \caption{RIS control block diagram.}
    \label{fig:RIS_control}
\end{figure}

Subsequently, we fabricated a $16\times16$ RIS as shown in Fig.~\ref{fig:RIS_board}. 
The RIS is comprised of four sub-arrays, each of which consists of $8\times8$ unit cells. 
Additionally, four identical control boards are designed and fabricated to independently control each unit cell of these sub-arrays. 
The block diagram of the control board is shown in Fig.~\ref{fig:RIS_control}. 
In each control board, we use a 8-bit shift register (SN74HC595), a 3-to-8 decoder (74HC238), and eight 8-bit D-type flip flops (74AC16374). 
A data acquisition equipment (DAQ) or a FPGA device is used to generate the ON/OFF signals of the unit cells. 
These ON/OFF signals are conveyed to the shift register, then forwarded to all D-type flip flops. 
Next, the flip flops load the ON/OFF signals when they are selected by the decoder. 
As a result, the desired state of every unit cell is loaded with the control signal. 
Finally, we combine the control board with the RIS as a sandwich structure. This facilitates extending the RIS to a larger scale. 
A LabVIEW code is programmed to properly control the board.

\subsection{Testbed Setup}
\label{subsec:testbed}
\begin{figure}[!htb]
    \centering
    \includegraphics[trim = 2cm 1cm 2cm 3cm, clip = true,width = 0.48\textwidth]{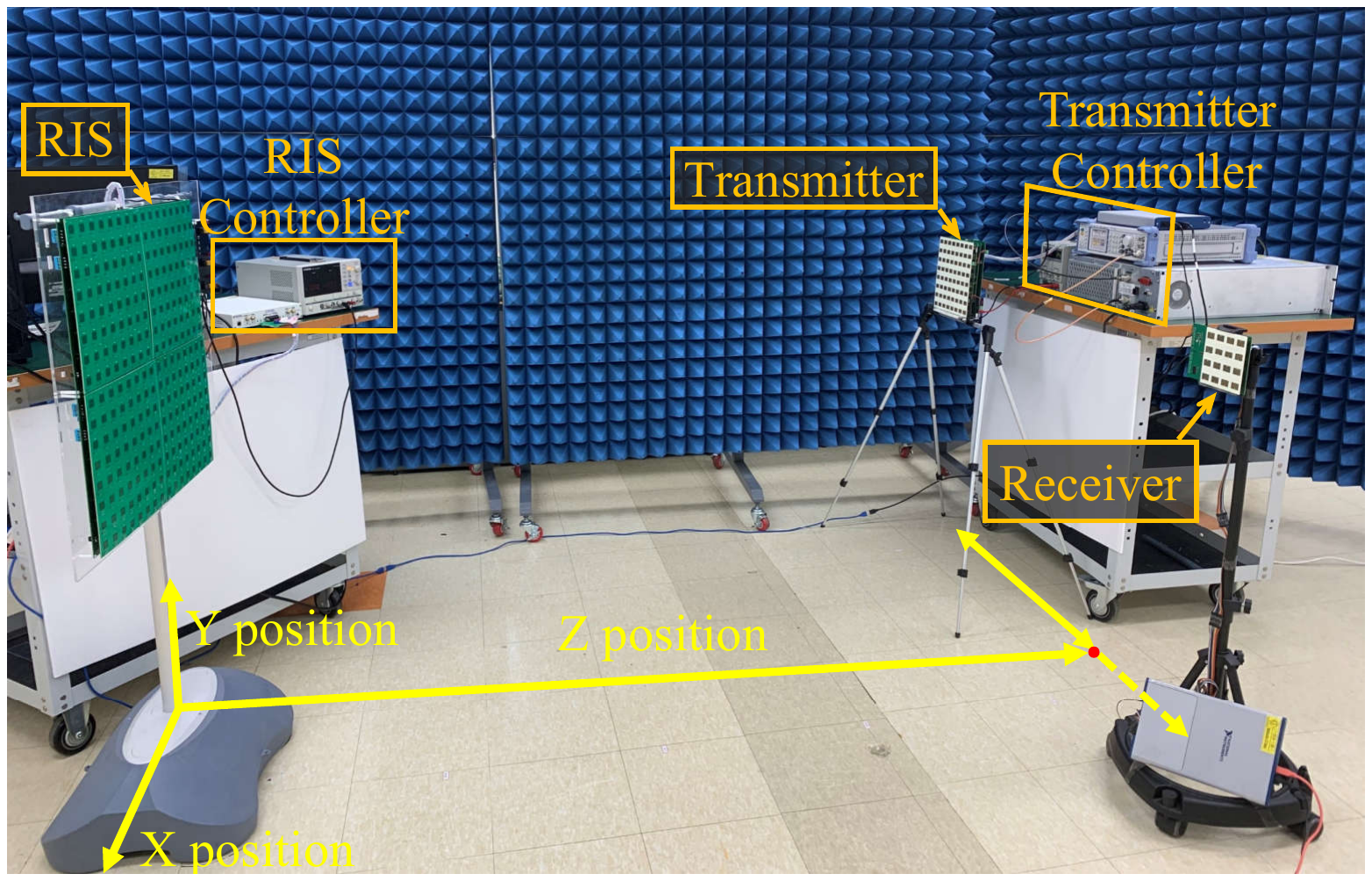}
    \caption{Experiment set-up.}
    \label{fig:testbed}
\end{figure}

The complete real-life testbed of RIS-aided WPT system is deployed as given in Fig.~\ref{fig:testbed}. Specifically, we aim to transfer the energy wirelessly from the transmitter to the receiver with relaying by the designed RIS. The experiment system consists of the fabricated RIS with its controller, a transmitter with its controller, and a receiver. An FPGA device (NI USRP 2944R) serves as the RIS controller. 

\begin{figure}[!htb]
    \centering
     \begin{subfigure}[c]{0.26\textwidth}
    \includegraphics[trim = 1cm 2cm 8cm 2cm,clip = true, width = \textwidth]{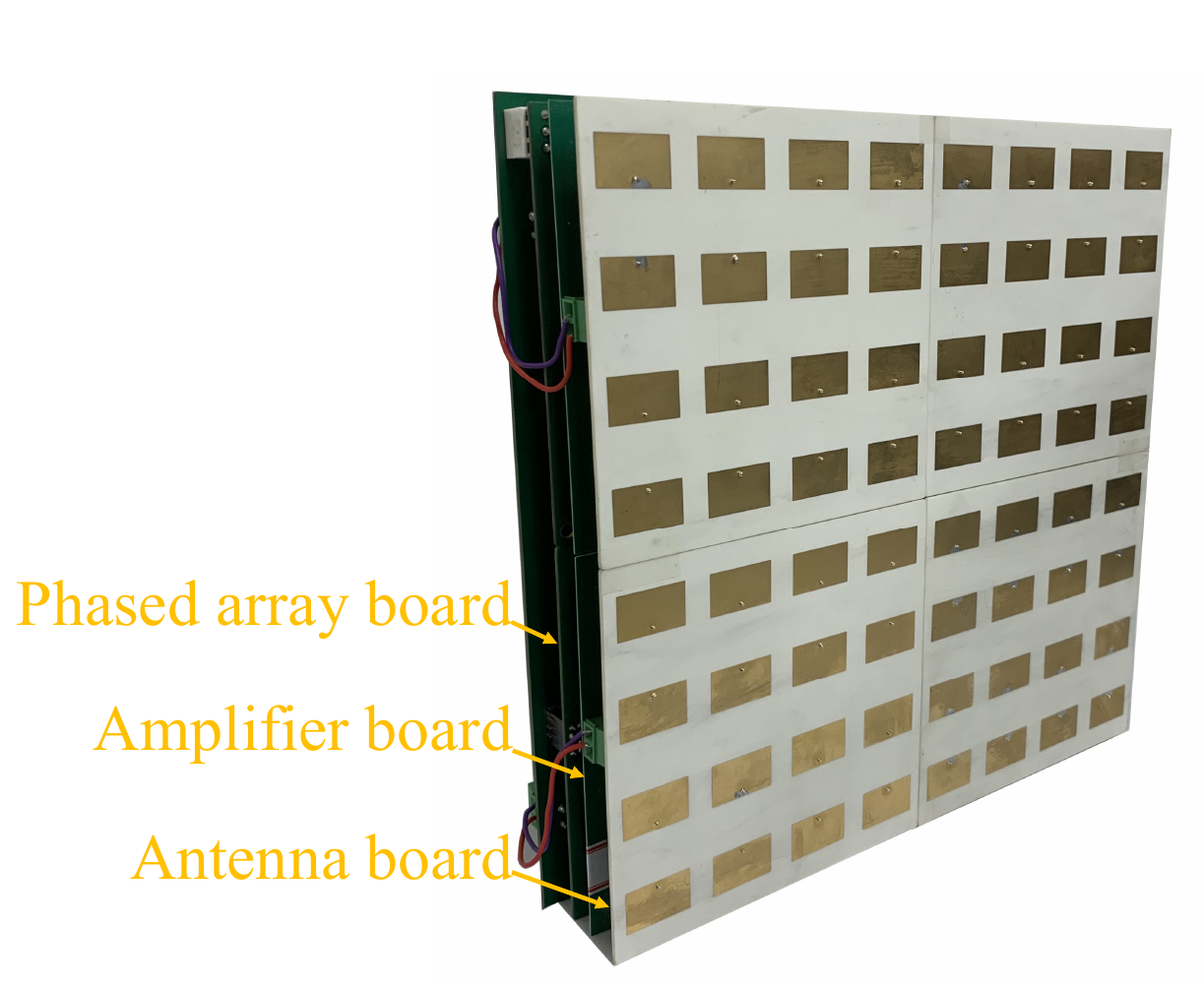}
    \caption{Transmitter board}
    \label{fig:tx_board}
    \end{subfigure}
    \hspace{-0.25cm}
    \begin{subfigure}[c]{0.23\textwidth}
    \includegraphics[trim = 5cm 2cm 5cm 2cm,clip = true, width = \textwidth]{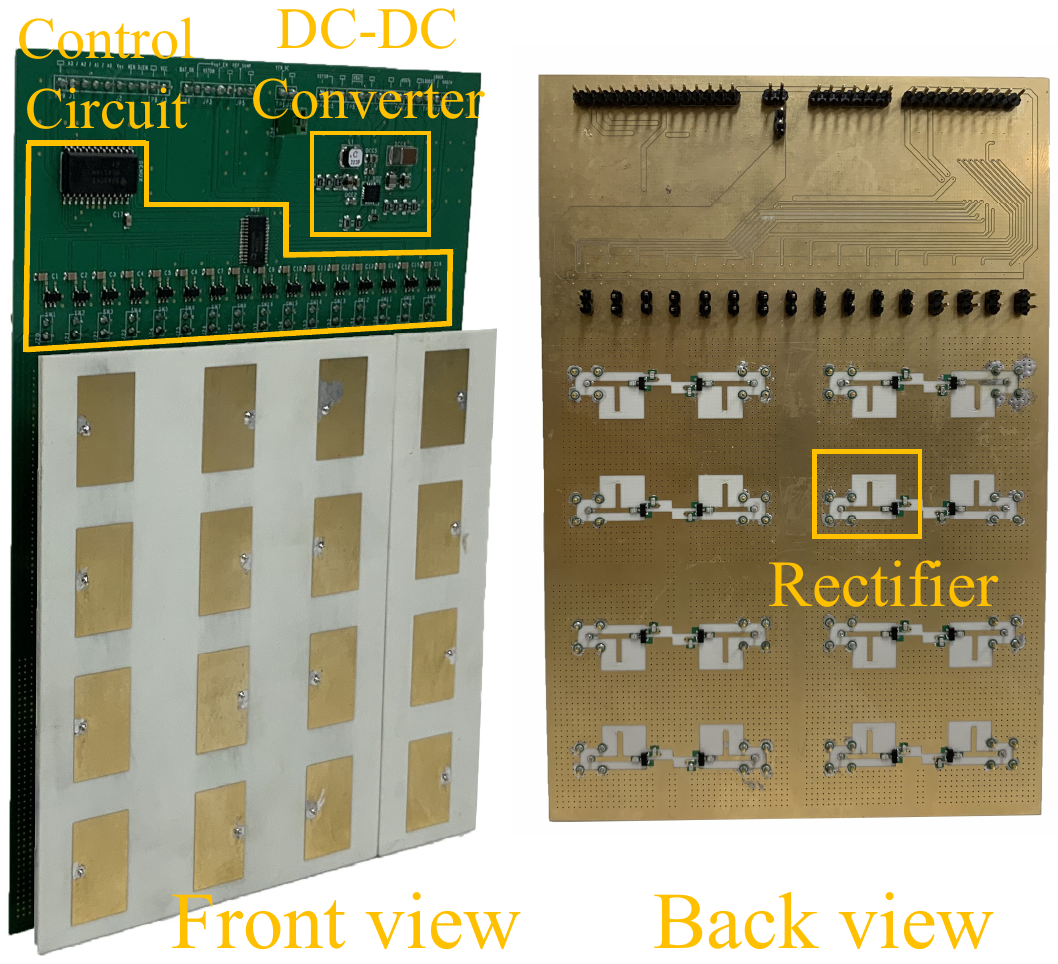}
    \caption{Receiver}
    \label{fig:rx_board}
    \end{subfigure}
    
    \caption{Transmitter and receiver prototype.}
    \label{fig:tx_rx_board}
\end{figure}

An $8\times8$ phased array antenna is used as a transmitter of the system. 
The transmitter array comprises four $4\times4$ unit modules which are made by stacking up a phased array board, an amplifier board, and an antenna board with a sandwich structure (Fig.~\ref{fig:tx_rx_board}(a)). 
We have designed and fabricated a supplemental board to combine four modules in one $8\times8$ transmitter. 
The board for combining modules consists of DC-DC converters, control signal lines, and two-stage Wilkinson dividers so that the transmitter can operate with only one DC input, one control signal input, and one RF input. 
Each RF path consists of a phase shifter, an RF switch, a power amplifier, and a shift register to control the phase shifter and the RF switch. 
A data acquisition equipment (NI USB-6351) is utilized to control the excitation phase of the transmitter. 
To provide the RF source to the transmitter, we employ a microwave signal generator (R\&S SMB 100A) in combination with an RF amplifier.

Fig.~\ref{fig:tx_rx_board}(b) shows $4\times4$ rectenna array receiver. 
To harvest the energy from the EM wave, multiple rectennas with high RF-to-DC conversion efficiency are integrated. 
Each rectifier is designed as a one-stage Dickson charge pump structure with one series pair diode and two capacitors.
RF switches controlled by a multiplexer and a decoder are connected to rectifiers.
By controlling these switches, we can measure the open-circuit voltage of each rectenna, which is an indicator of the receive power. 
The converted DC current from each rectenna is combined in parallel.

\section{Numerical Results}
\label{Sec:Results}
In this section, we present several simulation and experimental results for demonstrating the operation of the proposed algorithm.
\subsection{Simulation Results}
\label{subsec:simulresult}
To verify the proposed MTBS algorithm, we first show several simulation results with different RIS tile sizes. For testing the effectiveness of the proposed algorithm, we simulate the system with a larger RIS compared to the one in experiment. Specifically, we consider an 1-bit RIS with $40\times 40$ unit cells in this simulation. 
The RIS is divided into 400 tiles and 100 tiles with the corresponding RIS tile sizes of $2\times 2$ and $4\times 4$. The RIS is assumed to be located at the origin of the global Cartesian coordinate system. The phased array transmitter consists of $8\times8$ antenna elements and is positioned at (-0.5 m, 0 m, 2 m). The receiver is equipped with $4\times4$ antenna array and is located at (2 m, 1 m, 2 m). The antenna element $(2,2)$ is selected as the sensor antenna in the receiver. In the first iteration of the MTBS algorithm, the antenna element $(4,4)$ of the transmitter is turned on. In this simulation, the scanning beams over u-v coordinate of the transmitter/RIS tile are generated according to \eqref{eq:Txscanbeam} and \eqref{eq:RISscanbeam} given in Subsection \ref{subsec:scanningbeam}. Specifically, for a $M\times N$ planar array, we generate $2M\times2N$ scanning beams with $2M$ scanning points in the u-axis and $2N$ scanning points in the v-axis. Therefore, 256 scanning beams are generated for the $8\times8$ phased array transmitter. The number of scanning beams for RIS tile with $2\times2$ and $4\times4$ unit cells are 16 and 64, respectively.

\begin{figure}[!htb]
    \centering
     \begin{subfigure}[b]{0.5\textwidth}
    \includegraphics[trim = 4.5cm 2.5cm 5cm 3.5cm,clip = true, width = \textwidth]{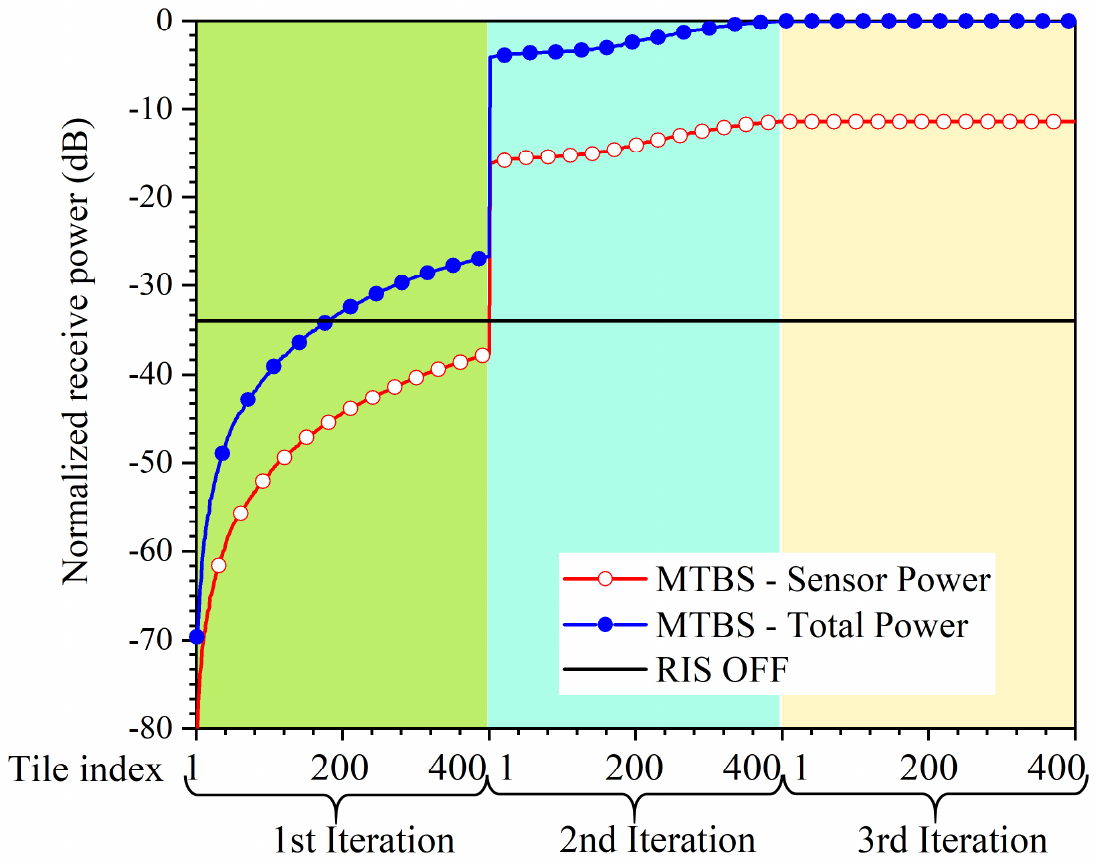}
    \caption{RIS tile size $2\times2$}
    \label{fig:Rxpow_22}
    \end{subfigure}
    \begin{subfigure}[b]{0.5\textwidth}
    \includegraphics[trim = 4.5cm 2.5cm 5cm 3.5cm,clip = true, width = \textwidth]{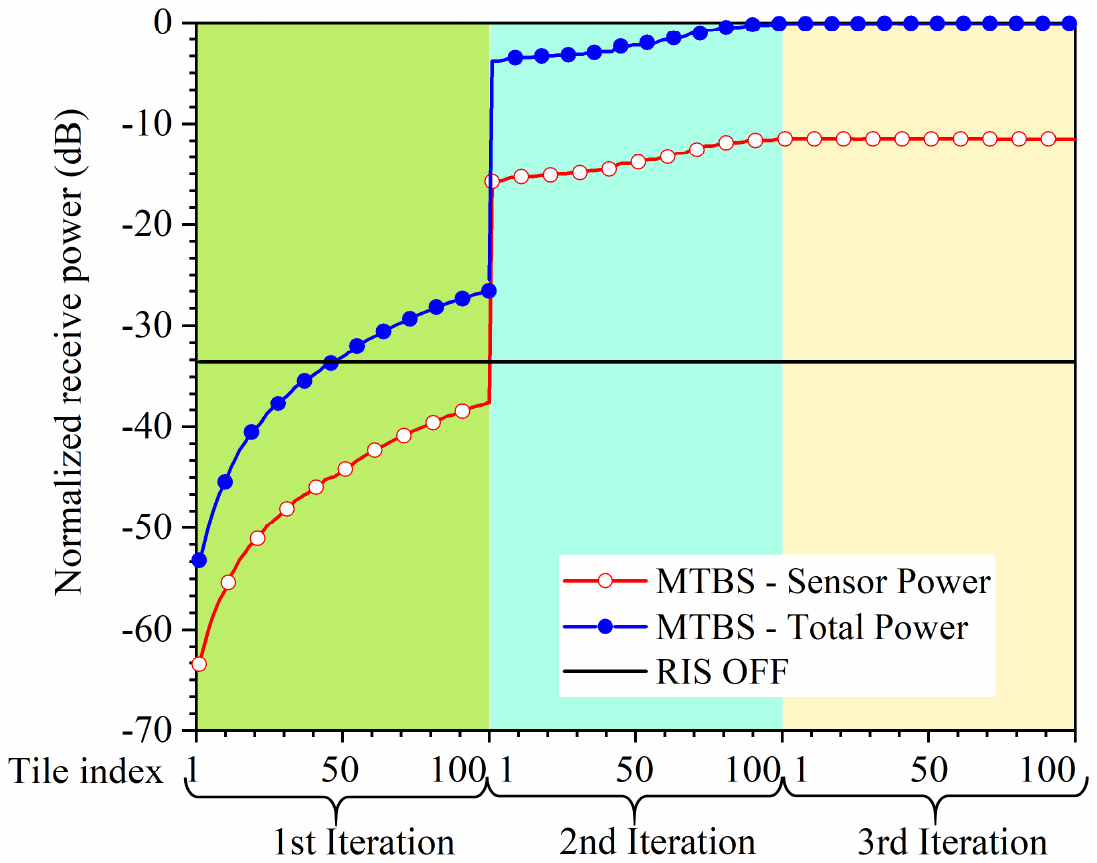}
    \caption{RIS tile size $4\times4$}
    \label{fig:Rxpow_44}
    \end{subfigure}
    \caption{Simulation receive power over scanning iterations.}
    \label{fig:simul_rxpow}
\end{figure}

Fig.~\ref{fig:simul_rxpow} shows the normalized receive power of the sensor antenna and the whole antenna array of the receiver over the scanning iterations with the RIS tile size of $2\times2$ and $4\times4$. The receive power of the sensor antenna and whole antenna array are indicated by ``MTBS - sensor power'' and ``MTBS - total power''. The ``RIS OFF'' represents the receive power when all unit cells of the RIS in OFF state. We can see that the RIS is effectively trained by the proposed algorithm. In the first iteration, the RIS tiles are well trained with about 45 dB improvement in the first case (i.e., RIS tile size of $2\times2$) and almost 28 dB gain in the second case (i.e., RIS tile size of $4\times4$) for both sensor antenna and whole antenna array. In the second scanning iteration, all antenna elements of the transmitter are turned on to transmit with the best beam. Therefore, there is a big jump in the receive power as shown in Fig.~\ref{fig:simul_rxpow} at the end of the first iteration. The RIS tiles are re-trained in the second iteration. There is around 5 dB improvement in the receive power that can be observed in both cases. The receive power in both cases remain stable in the third iteration. After executing the algorithm, about 35 dB gain in the total receive power (i.e., ``MTBS - total power'') is obtained compared to the ``RIS OFF'' case.
\begin{figure}[!htb]
    \centering
     \begin{subfigure}[b]{0.244\textwidth}
    \includegraphics[trim = 3.5cm 8.5cm 4cm 9.05cm,clip = true, width = \textwidth]{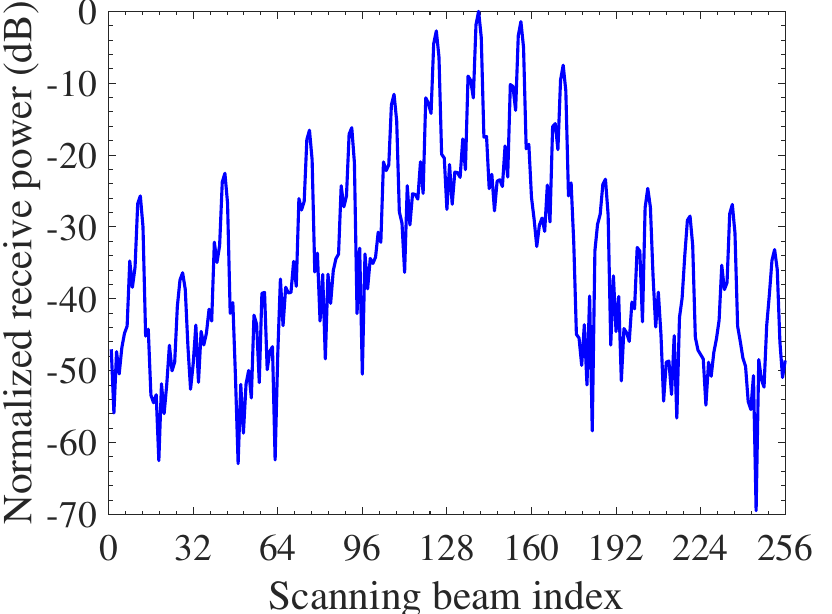}
    \caption{Receive power according to scanning beams}
    \label{fig:Txscan}
    \end{subfigure}
    \begin{subfigure}[b]{0.238\textwidth}
    \includegraphics[trim = 3cm 7.75cm 3.5cm 8cm,clip = true, width = \textwidth]{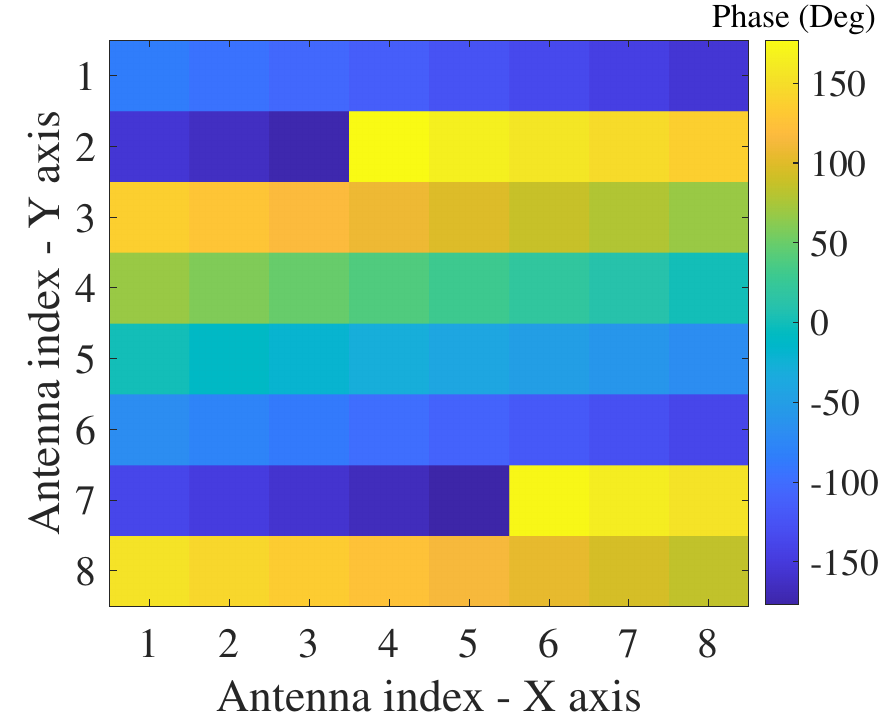}
    \caption{Transmitter optimal phase excitation}
    \label{fig:Txpha}
    \end{subfigure}
    \caption{Transmitter scanning process.}
    \label{fig:simul_tx}
\end{figure}

\begin{figure}[!htb]
    \centering
     \begin{subfigure}[b]{0.24\textwidth}
    \includegraphics[trim = 5.5cm 1.25cm 5cm 0.5cm,clip = true, width = \textwidth]{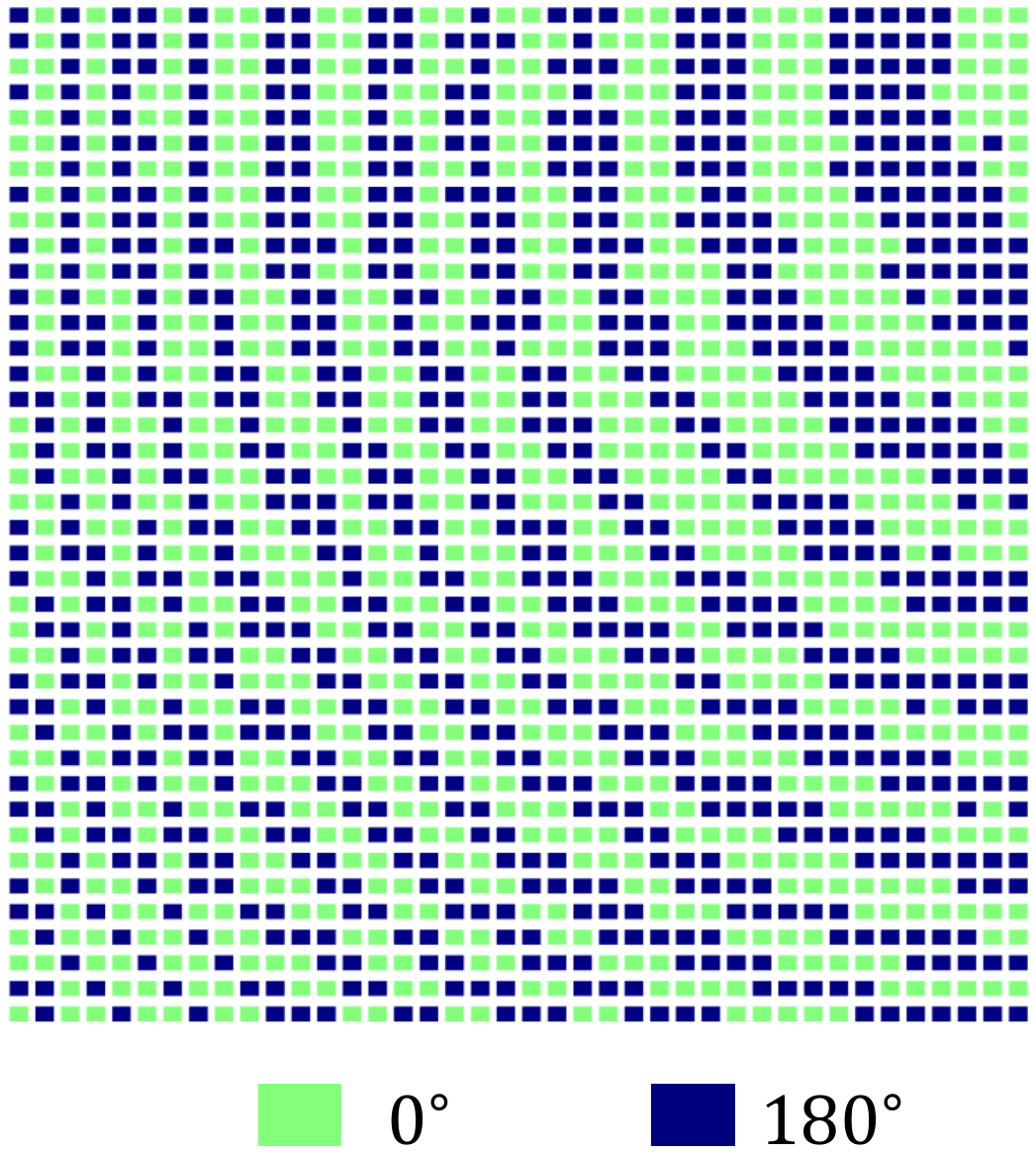}
    \caption{RIS tile size $2\times2$}
    \label{fig:RISpat_22}
    \end{subfigure}
    \begin{subfigure}[b]{0.24\textwidth}
    \includegraphics[trim = 5.5cm 1.25cm 5cm 0.5cm,clip = true, width = \textwidth]{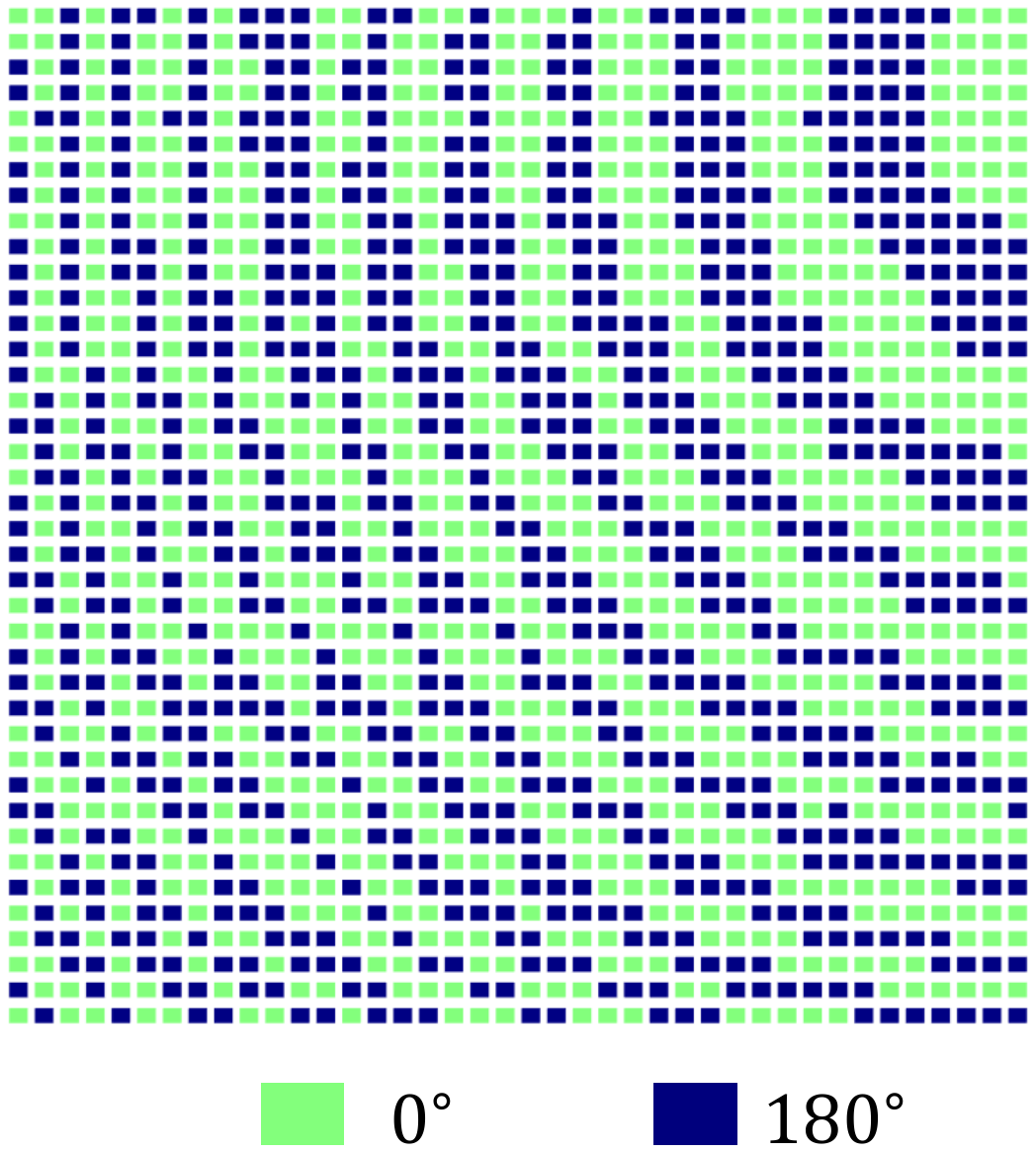}
    \caption{RIS tile size $4\times4$}
    \label{fig:RISpat_44}
    \end{subfigure}
    
    
    \caption{RIS optimal phase distribution.}
    \label{fig:simul_pat}
\end{figure}

The transmitter scanning results in the first iteration are presented in Fig.~\ref{fig:simul_tx}. The normalized receive power corresponding to each scanning beam is given in Fig.~\ref{fig:simul_tx}(a). It is noted that the 140th scanning beam results in the highest receive power. Fig.~\ref{fig:simul_tx}(b) indicates the corresponding phase distribution of that scanning beam. The RIS optimal phase distributions of the two cases are given in Fig.~\ref{fig:simul_pat}. We can see that a convex lens-like pattern is formed in both cases to focus the power beam to the receiver. This cannot be done by a conventional far-field scanning algorithm.

\subsection{Experimental Results}
\label{subsec:experiment}
In this subsection, we present experimental results obtained by performing the wireless power transfer experiments in the real test scenarios with the fabricated RIS.

\subsubsection{Algorithm Validation}
\label{subsubsec:validation}

\begin{figure}[!htb]
    \centering
    \includegraphics[trim = 4cm 2.5cm 3cm 2.5cm, clip = true,width = 0.48\textwidth]{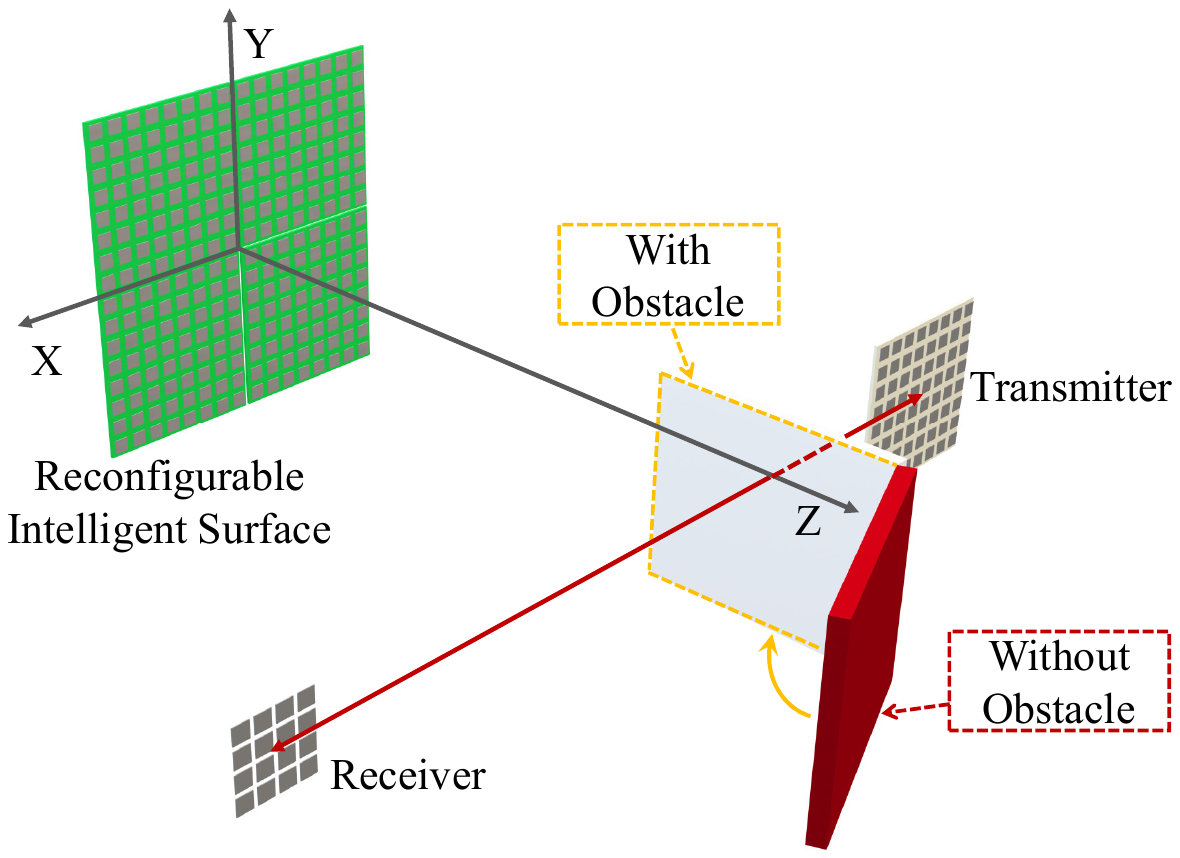}
    \caption{Testbed coordinate system.}
    \label{fig:testbedcoord}
\end{figure}

\begin{figure}[!htb]
    \centering
    \includegraphics[trim = 4.5cm 2.5cm 6cm 3.5cm, clip = true,width = 0.48\textwidth]{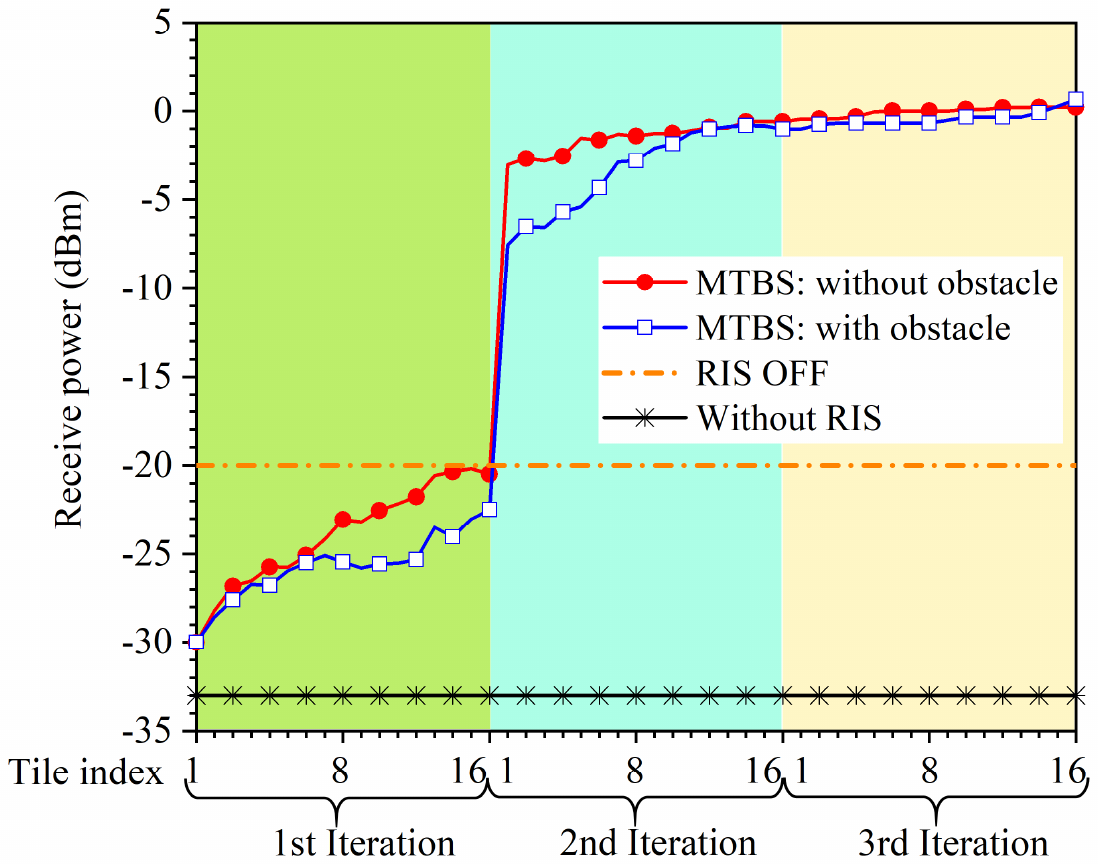}
    \caption{Receive power over the iteration (RIS tile size $4\times4$).}
    \label{fig:exp_rx_44}
\end{figure}

\begin{figure}[!htb]
    \centering
     \begin{subfigure}[b]{0.45\textwidth}
    \includegraphics[trim = 3cm 8.5cm 3cm 9cm,clip = true, width = \textwidth]{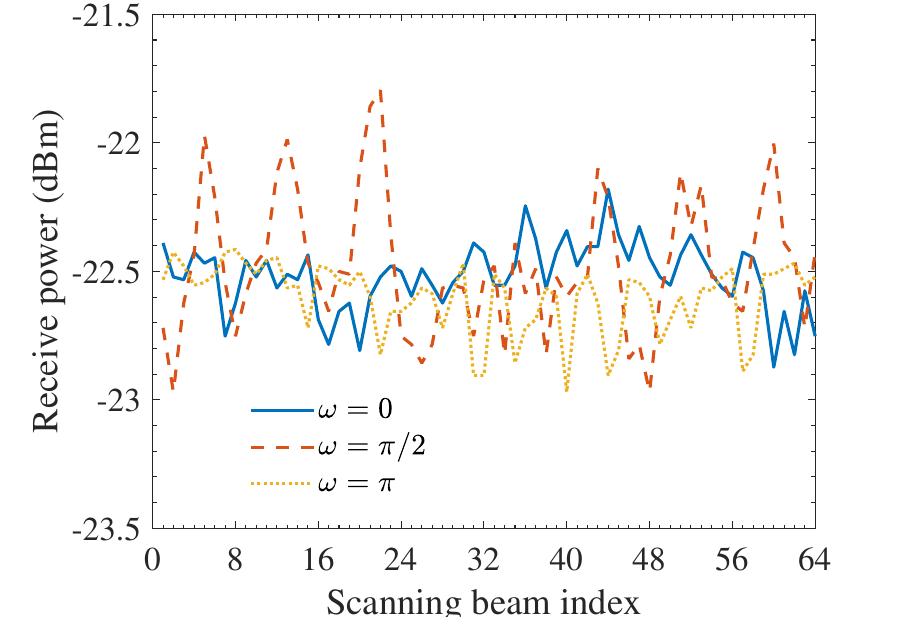}
    \caption{Receive power with different phase control}
    \label{fig:RISscan1}
    \end{subfigure}
    \begin{subfigure}[b]{0.45\textwidth}
    \includegraphics[trim = 1cm 0.5cm 1cm 1.25cm,clip = true, width = \textwidth]{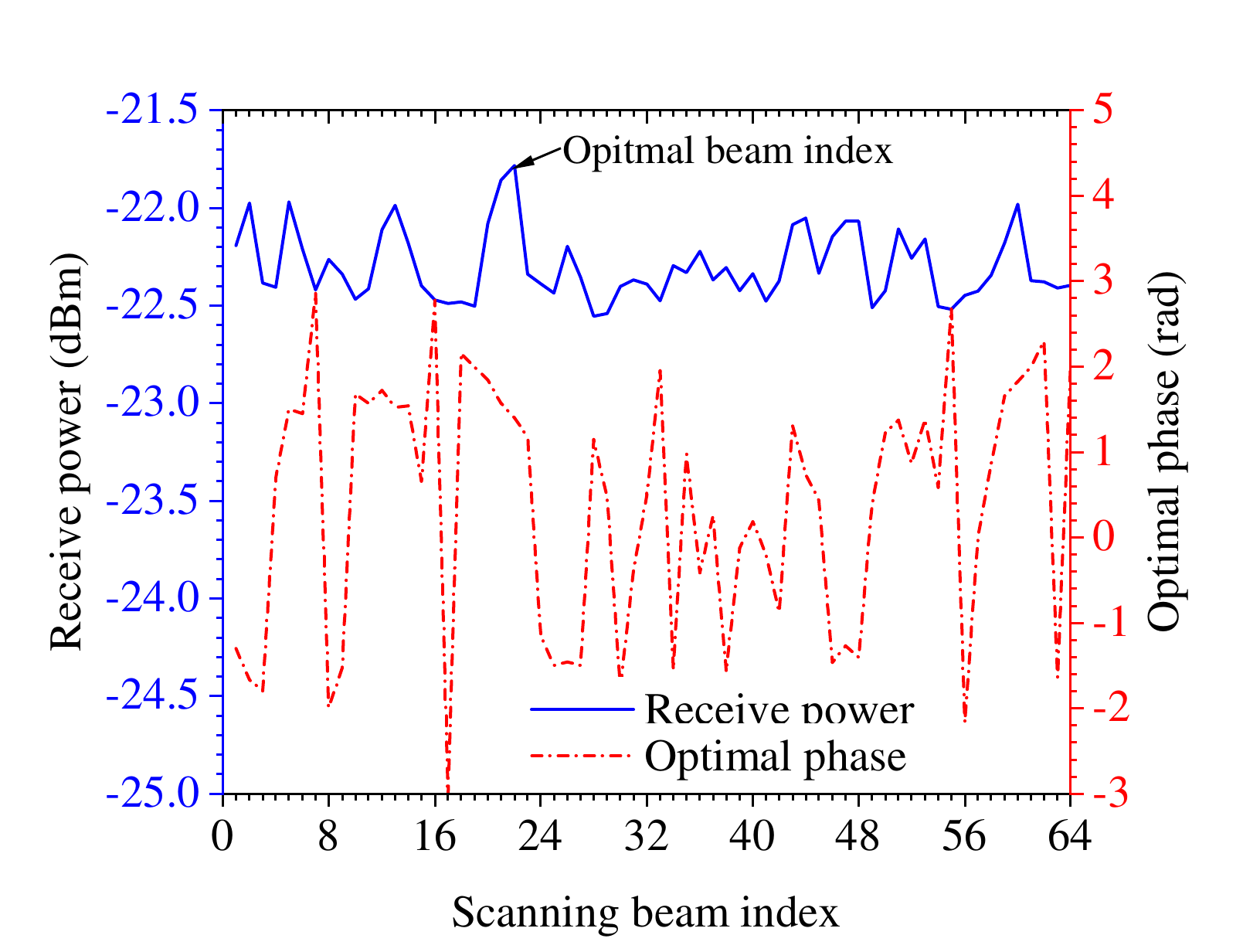}
    \caption{Optimal phase control and corresponding receive power}
    \label{fig:RISscan2}
    \end{subfigure}
    \caption{RIS tile scanning process in experiments.}
    \label{fig:RISscan}
\end{figure}

\begin{figure}[!htb]
    \centering
     \begin{subfigure}[b]{0.235\textwidth}
    \includegraphics[trim = 4cm 8.75cm 4.5cm 9.25cm,clip = true, width = \textwidth]{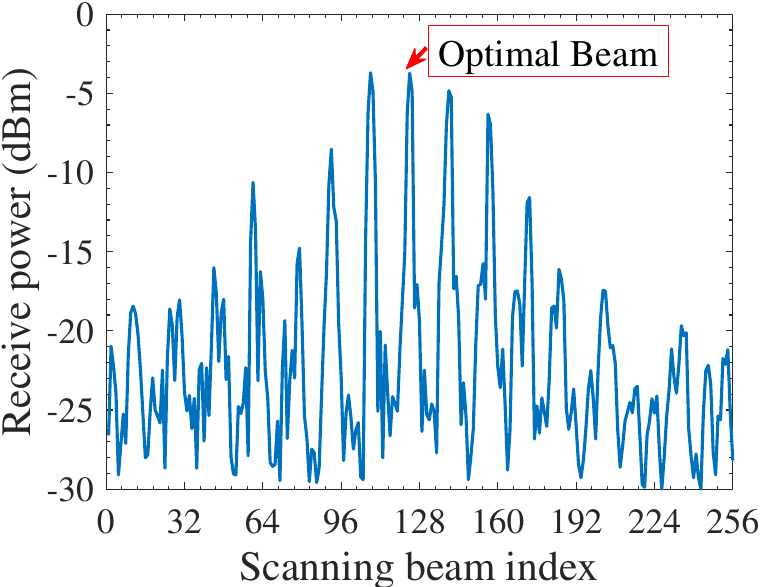}
    \caption{Receive power according to scanning beam}
    \label{fig:ExpTxscan}
    \end{subfigure}
    \begin{subfigure}[b]{0.245\textwidth}
    \includegraphics[trim = 3.75cm 8.5cm 3.5cm 9.1cm,clip = true, width = \textwidth]{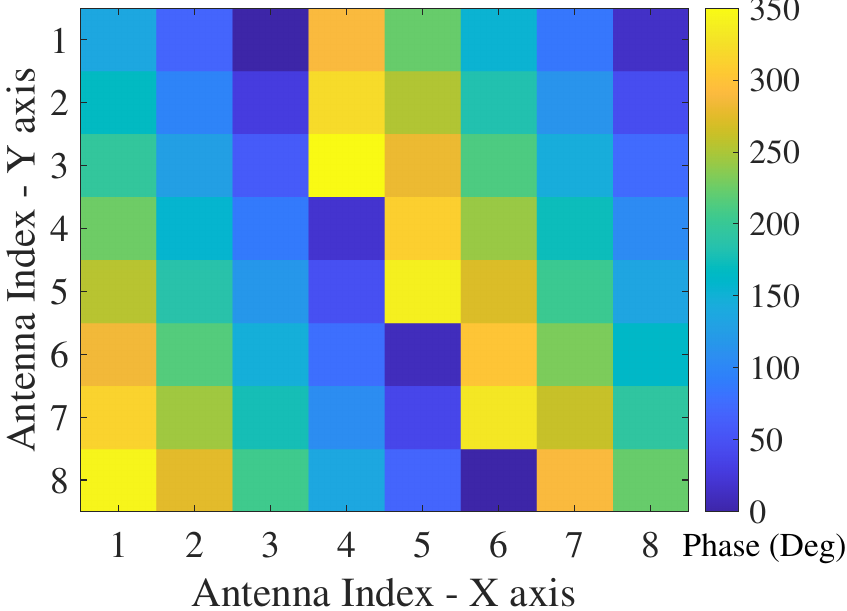}
    \caption{Transmitter optimal phase control}
    \label{fig:ExpTxpha}
    \end{subfigure}
    \caption{Transmitter scanning process in experiment.}
    \label{fig:Exp_tx}
\end{figure}

\begin{figure}[!htb]
    \centering
     \begin{subfigure}[b]{0.22\textwidth}
    \includegraphics[trim = 4.5cm 9cm 6.5cm 9.25cm,clip = true, width = \textwidth]{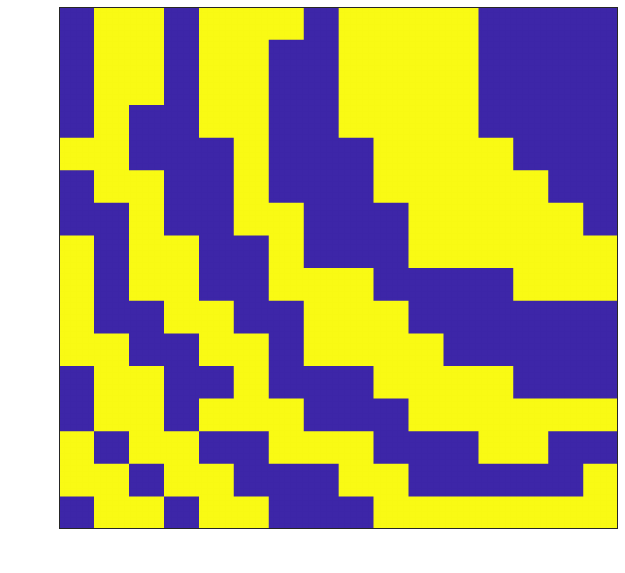}
    \caption{Without obstacle}
    \label{fig:ExpRISpat4}
    \end{subfigure}
    \begin{subfigure}[b]{0.26\textwidth}
    \includegraphics[trim = 4.5cm 9cm 4.5cm 8.75cm,clip = true, width = \textwidth]{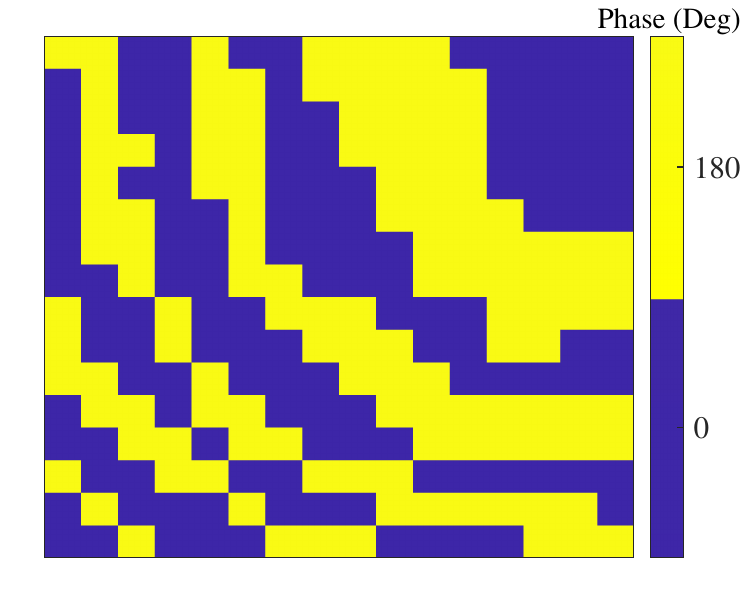}
    \caption{With obstacle}
    \label{fig:ExpRISpat4ob}
    \end{subfigure}
    \caption{RIS optimal phase distribution (RIS tile size $4\times4$).}
    \label{fig:ExpRISpat44}
\end{figure}

\begin{figure}[!htb]
    \centering
    \includegraphics[trim = 4.5cm 2.5cm 6.5cm 3.5cm,clip = true, width = 0.45\textwidth]{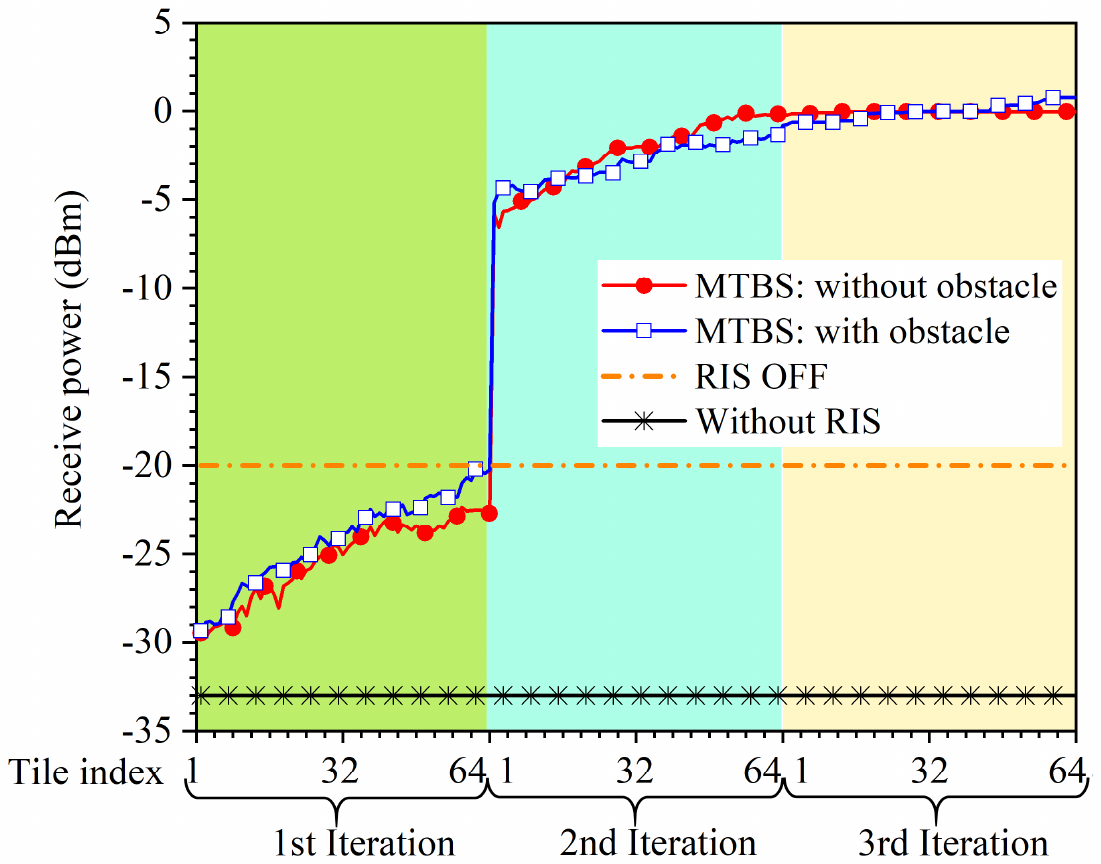}
    \caption{Receive power over the iteration (RIS tile $2\times2$).}
    \label{fig:Exp_rx_22}
\end{figure}

\begin{figure}[!htb]
    \centering
     \begin{subfigure}[b]{0.22\textwidth}
    \includegraphics[trim = 4.5cm 9cm 6.7cm 9.25cm,clip = true, width = \textwidth]{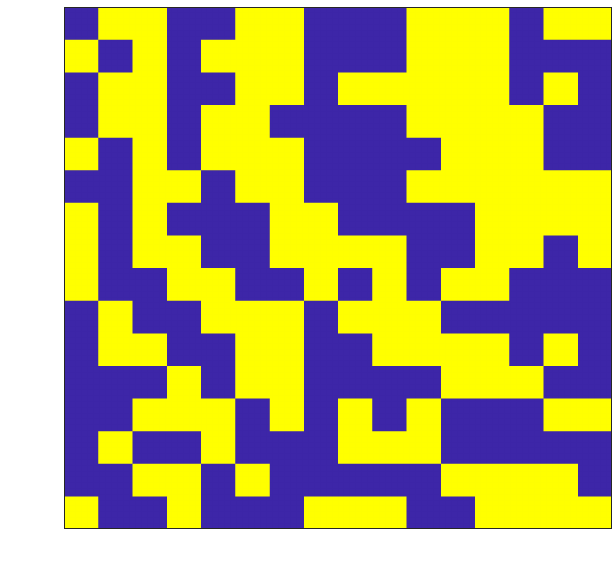}
    \caption{Without obstacle}
    \label{fig:ExpRISpat2}
    \end{subfigure}
    \begin{subfigure}[b]{0.26\textwidth}
    \includegraphics[trim = 4.5cm 9cm 4.75cm 8.75cm,clip = true, width = \textwidth]{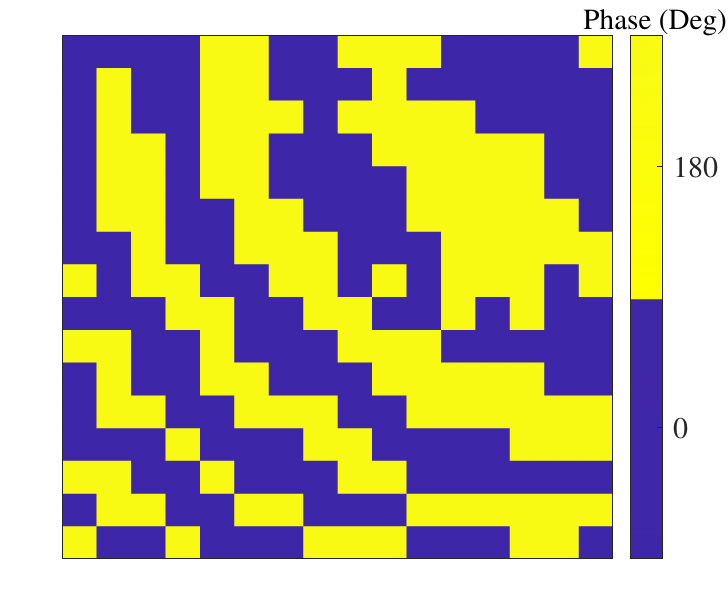}
    \caption{With obstacle}
    \label{fig:ExpRISpat2ob}
    \end{subfigure}
    \caption{RIS optimal phase distribution (RIS tile size $2\times2$).}
    \label{fig:ExpRISpat22}
\end{figure}

\begin{figure}[!htb]
    \centering
    \includegraphics[trim = 4.5cm 2.5cm 6.5cm 3.5cm,clip = true, width = 0.45\textwidth]{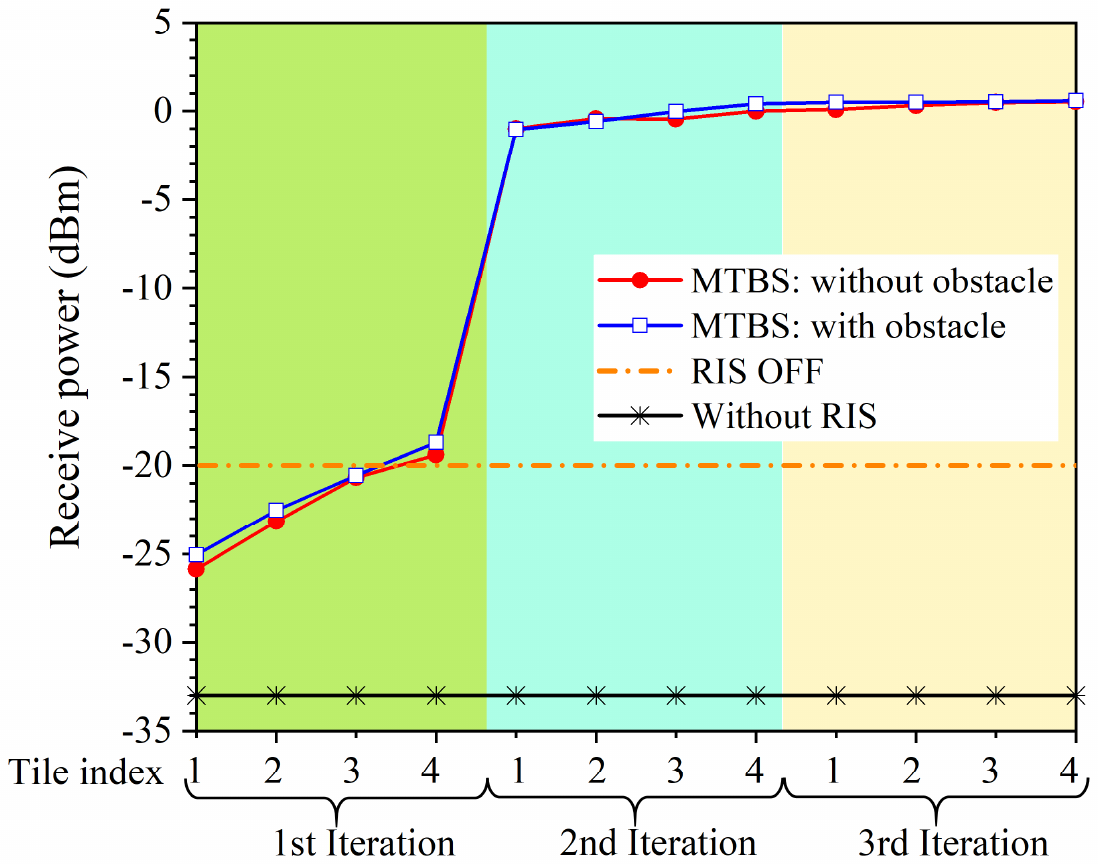}
    \caption{Receive power over the iteration (RIS tile size $8\times8$).}
    \label{fig:Exp_rx_88}
\end{figure}

We first verify the beam focusing capability of the proposed system. The coordinate system of the testbed in Fig.~\ref{fig:testbed} is given in Fig.~\ref{fig:testbedcoord}. The RIS is set at the origin of the coordinate system. In this test scenario, the transmitter is located at (-0.5 m, 0 m, 1.5 m), and the receiver is placed at (1.5 m, 0.5 m, 2 m). Similar to the simulation, the scanning beams over u-v coordinate of the transmitter/RIS tile are generated in the same way. Thus, 256 scanning beams are used for the $8\times8$ phased array transmitter. The number of scanning beams for RIS tile sizes of $2\times2$, $4\times4$, and $8\times8$ are 16, 64, and 256, respectively.
In the first iteration, one antenna element of the transmitter (i.e., antenna element $(4,4)$) is turned on and transmits with the power of 23 dBm. In the next iteration, the whole phased array transmitter is active and the transmit power of each element is 15 dBm. Higher transmit power in the first iteration ensures a sufficiently high receive power for operating the algorithm correctly. The receiver measures the receive power at the sensor antenna which is the antenna element $(2,2)$. It is noted that the receive power of Figs.~\ref{fig:exp_rx_44}--\ref{fig:Exp_rx_88} is the receive RF power measured by a single sensor antenna at the receiver. 

To demonstrate the benefit of the RIS, we insert the obstacle between the transmitter and receiver to block the direct channel (see Fig.~\ref{fig:testbedcoord}). The receive power for this case is presented by ``MTBS: with obstacle'', and the one without the obstacle corresponds to ``MTBS: without obstacle''.  In addition, after executing the proposed algorithm, we remove the RIS to clearly show the RIS effect. The corresponding result is indicated by ``Without RIS''.

Fig.~\ref{fig:exp_rx_44} presents the receive power of the sensor antenna  over the iterations when the RIS tile size $4\times4$ is used in the scanning.
We can observe from Fig.~\ref{fig:exp_rx_44} that very good measured results are obtained. After the first iteration, while around 9 dB improvement is seen in the case without obstacle, 7 dB gain is achieved when an obstacle is inserted. In the next scanning iteration, we can see a great enhancement of the receive power in the latter case (i.e. MTBS: with obstacle). Subsequently, the same performance is obtained for the two cases after executing the algorithm. It is clear that a fraction of the transmit power to some RIS tiles is blocked by the obstacle when one antenna element is active. As a result, these RIS tiles are not well trained in the first iteration. Hence, better optimal control parameters for the RIS tiles are updated in the next iteration. Finally, we can see that around 20 dB gain in the receive power is achieved with the proposed algorithm in comparison with the ``RIS OFF'' case. The receive power degrades almost 35 dB when the RIS is removed (i.e., ``Without RIS'' case).

The measured data while scanning the 12th RIS tile in the first iteration of ``MTBS: without obstacle'' case is shown in Fig.~\ref{fig:RISscan}(a). The receive power with respect to three different phases of each scanning beam is measured. Using these measured data, the optimal phase control of each scanning beam is calculated by using \eqref{eq:omegaopt2}. According to \eqref{eq:calpow}, the corresponding receive power is obtained. The calculated optimal phase control and the corresponding receive power for all scanning beams are given in Fig.~\ref{fig:RISscan}(b). We can observe that the 22th scanning beam is the optimal one with the highest receive power of about -21.7 dBm.

Fig.~\ref{fig:Exp_tx}(a) presents the receive power with respect to the scanning beams when the transmitter is scanned in the first iteration. We can see that the highest receive power is around -4 dBm at the 124th scanning beam. This scanning beam is selected as the optimal beam for transmitting in the next iteration. The corresponding phase distribution of the optimal beam of the transmitter is shown in Fig.~\ref{fig:Exp_tx}(b).

The optimal phase distribution of the RIS for the two cases of with and without obstacles are given in Fig.~\ref{fig:ExpRISpat44}. A very good agreement between the two cases confirms the same receive power given in Fig.~\ref{fig:exp_rx_44}. Up to this point, we can observe that the proposed algorithm operates effectively even an obstacle inserted between the transmitter and the receiver. It proves that by deploying the proposed algorithm, RIS can greatly enhance the receive power of WPT systems even when the direct channel deteriorates.

\begin{figure}[!htb]
    \centering
     \begin{subfigure}[b]{0.22\textwidth}
    \includegraphics[trim = 4.5cm 9cm 6.5cm 9.25cm,clip = true, width = \textwidth]{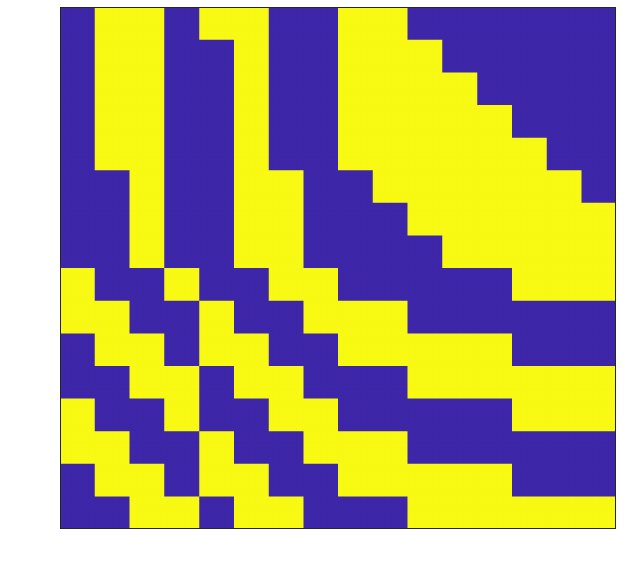}
    \caption{Without obstacle}
    \label{fig:ExpRISpat8}
    \end{subfigure}
    \begin{subfigure}[b]{0.26\textwidth}
    \includegraphics[trim = 4.5cm 9cm 4.5cm 8.75cm,clip = true, width = \textwidth]{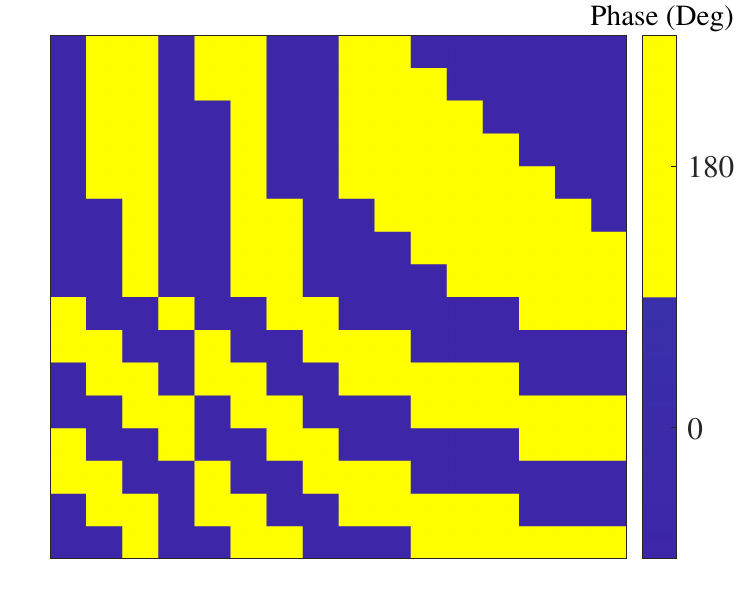}
    \caption{With obstacle}
    \label{fig:ExpRISpat8ob}
    \end{subfigure}
    \caption{RIS optimal phase distribution (RIS tile size $8\times8$).}
    \label{fig:ExpRISpat88}
\end{figure}

Furthermore, the experiments with RIS tile sizes of $2\times2$ and $8\times8$ are conducted and the results are presented in Figs.~\ref{fig:Exp_rx_22}, \ref{fig:ExpRISpat22}, \ref{fig:Exp_rx_88}, and \ref{fig:ExpRISpat88}. We can see that very good results are obtained. The almost same performance is given between two cases (with and without obstacle) when scanning with both RIS tile sizes. There are around 20 dB and 35 dB gain of the receive power in comparison with the ``RIS OFF'' case and ``Without RIS'' case, respectively. It can be seen from Fig.~\ref{fig:ExpRISpat22} that some RIS tiles are not very well trained. This is because the RIS tile size (i.e., $2\times2$) is relatively small, which leads to small receive power causing the error in calculating the optimal parameters. However, a convex lens-like phase distribution can still be formed, and a good result is achieved after the MTBS algorithm is executed. In addition, a very good match between the two results of the RIS tile size $8\times8$ case can be observed in Fig.~\ref{fig:ExpRISpat88}(a) and Fig.~\ref{fig:ExpRISpat88}(b). This explains the similar trend over scanning iterations between ``MTBS: with obstacle'' and ``MTBS: without obstacle'' presented in Fig.~\ref{fig:Exp_rx_88}. 
\subsubsection{Power Transfer Experiment}
\begin{figure}[!htb]
    \centering
    \includegraphics[trim = 2cm 2cm 0cm 1cm,clip = true, width = 0.48\textwidth]{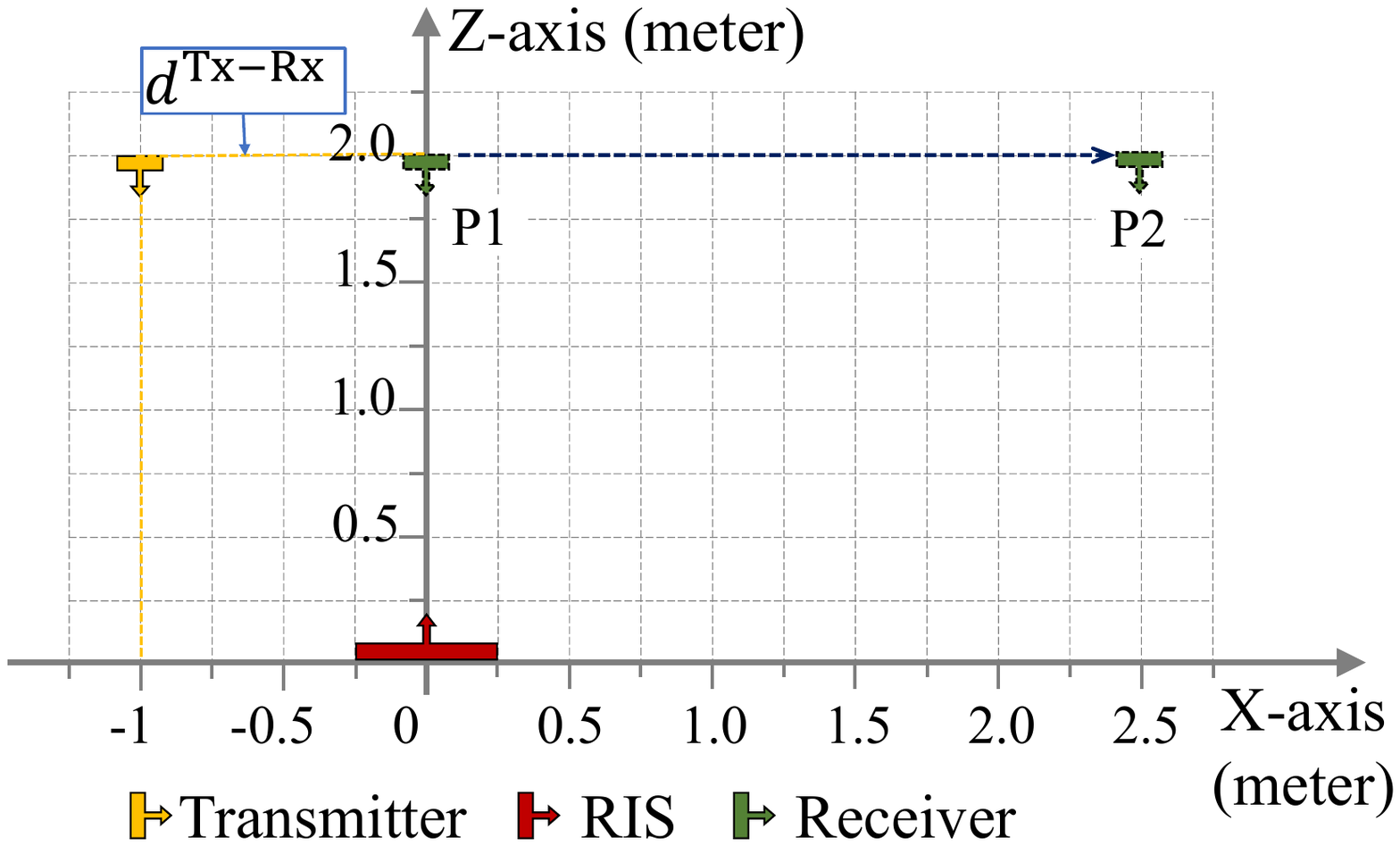}
    \caption{Transmitter, receiver, and RIS position.}
    \label{fig:Exptestpos}
\end{figure}

Finally, we perform wireless power transfer with the fabricated testbed and measure the power transfer efficiency over the distances. In this test scenario, the transmitter and receiver are located 2 meters away from the RIS in the z-direction. The transmitter, receiver, and RIS are set at the same height. Fig.~\ref{fig:Exptestpos} shows the positions of the transmitter, receiver, and RIS in x-z plane.

\begin{figure}[!htb]
    \centering
    \includegraphics[trim = 2cm 0.5cm 1cm 2cm,clip = true, width = 0.48\textwidth]{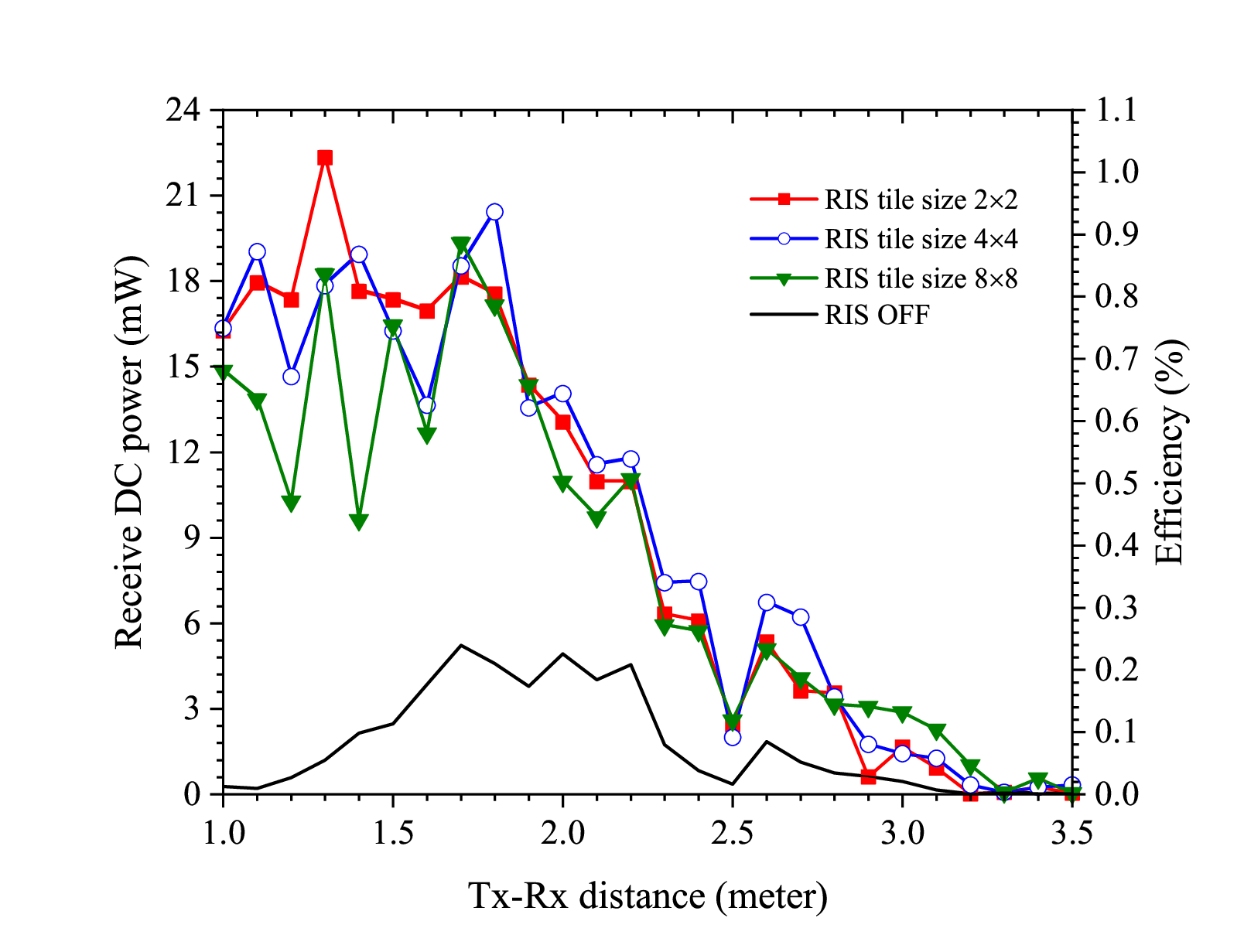}
    \caption{Receive power and power transfer efficiency over Tx-Rx distance.}
    \label{fig:Exppoweff}
\end{figure}

Specifically, the RIS is located at the origin of the coordinate system. The transmitter is positioned at (-1 m, 2 m), and the receiver is at ($x$, 2 m) with $x$ ranging from 0 m to 2.5 m. The distance between the transmitter and the receiver is indicated by $d^\text{Tx-Rx}$ in Fig.~\ref{fig:Exptestpos}. Henceforth, we will call this distance the Tx-Rx distance. Indeed, the receiver is moved along with the x-axis from position P1 to position P2 to vary Tx-Rx distance from 1 meter to 3.5 meters. It is noted that the obstacle is removed for this test. Furthermore, for a real WPT, the RF signal is rectified and stored in the energy storage device (e.g., super-capacitor). Therefore, in this test the rectified DC power of the whole antenna array is measured as the receive DC power. It is different from the RF receive power from the sensor antenna presented in the previous results. The total transmit power is 2.1 W. The corresponding receive DC power of the receiver according to the Tx-Rx distance is recorded, and the power transfer efficiency is calculated accordingly. In this experiment, the results with three different RIS tile sizes are obtained and compared with that of ``RIS OFF'' case. Fig.~\ref{fig:Exppoweff} shows the obtained results (i.e., receive DC power and power transfer efficiency) according to the Tx-Rx distance.

Fig.~\ref{fig:Exppoweff} demonstrates that for all RIS tile-size cases, the receive DC power fluctuates around 18 mW while varying the Tx-Rx distance from 1 m to 1.9 m before dropping to 2 mW at 2.5 m. It is because the steering angle ($\phi^{\text{RIS-Rx}}_k$) gets wider when the test distance increases. When it approaches the steering limit of the RIS, a part of the reflected power from the RIS is directed to the other side of the steering angle. It is similar to the grating lobe phenomenon when steering the beam in the planar phased array antenna. Hence, the performance deteriorates. Moreover, we can notice that the receive DC power of the ``RIS OFF'' case increases to nearly 5 dBm as the Tx-Rx distance approaches 2 m. This is due to the specular reflection appearing at that point. The receive DC power for that case degrades instantly when the receiver continues moving along with x-axis.

It is observed that around 22.5 mW is transferred to the receiver at 1.3 m Tx-Rx distance when the RIS tile is scanned with the RIS tile size of $2\times2$. It corresponds to almost 1.05\% power transfer efficiency. The reason for this low efficiency is that the actual channel distance (i.e., transmitter-RIS-receiver) is almost 4.25 m. This efficiency is comparable with that presented in \cite{ref23}. We can expect that higher power transfer efficiency can be achieved by deploying a larger RIS system. However, this receive DC power is sufficient for charging the low-power IoT receiver and keeping it alive \cite{ref24}. 

\begin{figure}[!htb]
    \centering
     \begin{subfigure}[b]{0.5\textwidth}
    \includegraphics[trim = 1cm 7cm 1cm 4.5cm,clip = true, width = \textwidth]{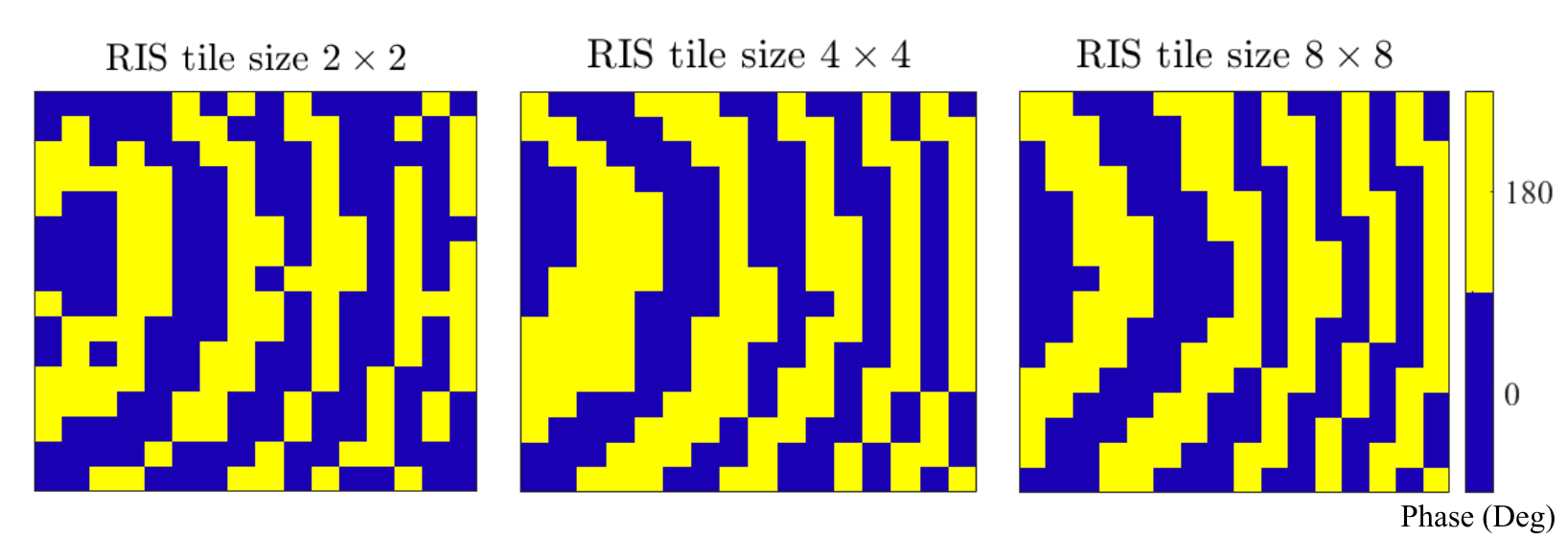}
    \caption{Tx-Rx distance = 1 m}
    \label{fig:expat0m}
    \end{subfigure}
    \begin{subfigure}[b]{0.5\textwidth}
    \includegraphics[trim = 1cm 7cm 1cm 4.5cm,clip = true, width = \textwidth]{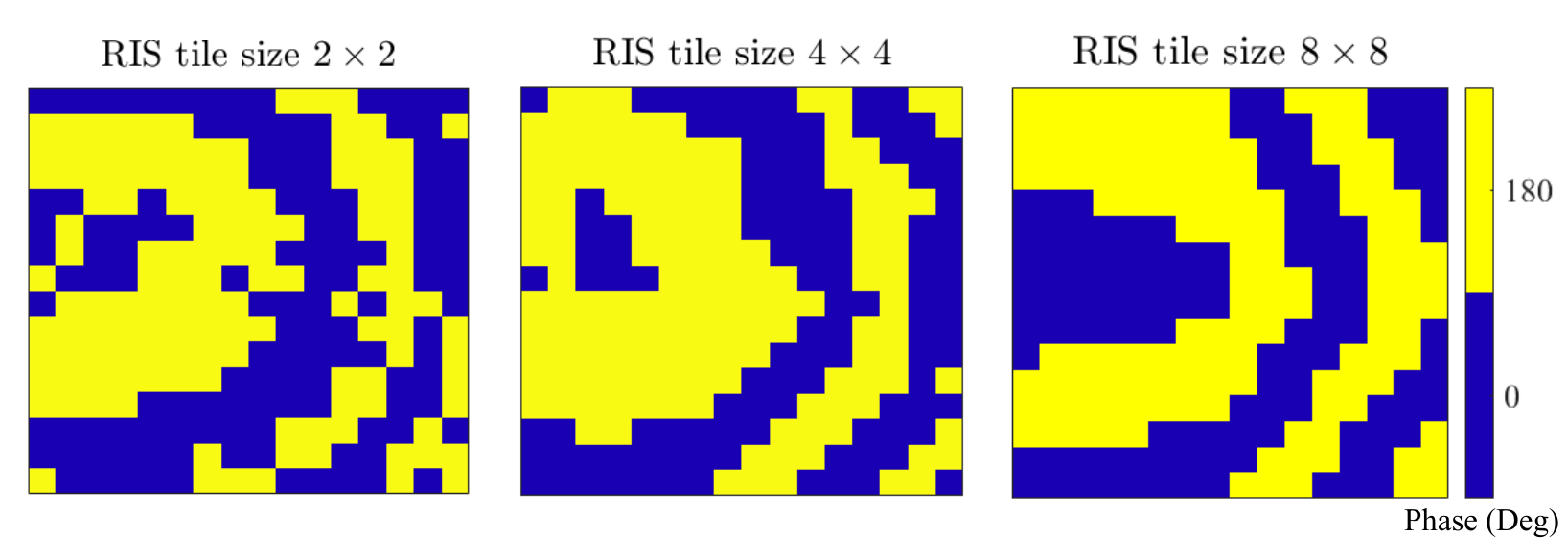}
    \caption{Tx-Rx distance = 1.5 m}
    \label{fig:exppat0.5m}
    \end{subfigure}
     \begin{subfigure}[b]{0.5\textwidth}
    \includegraphics[trim = 1cm 7cm 1cm 4.5cm,clip = true, width = \textwidth]{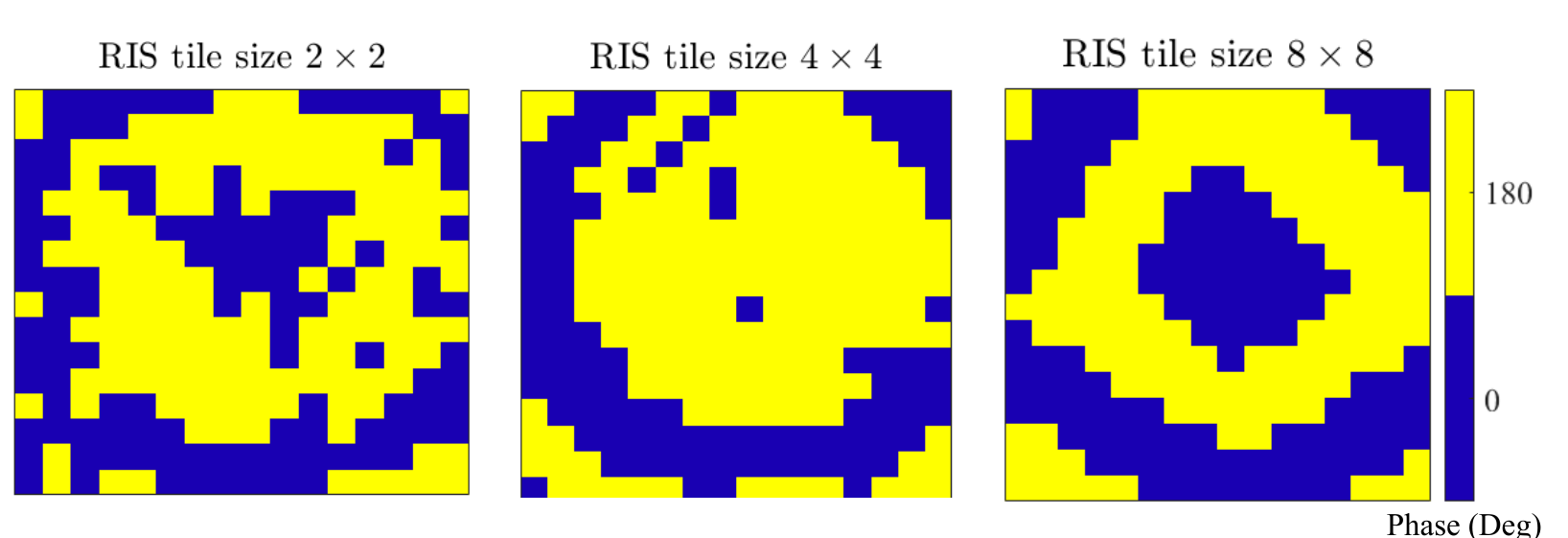}
    \caption{Tx-Rx distance = 2 m}
    \label{fig:expat1m}
    \end{subfigure}
    \begin{subfigure}[b]{0.5\textwidth}
    \includegraphics[trim = 1cm 7cm 1cm 4.5cm,clip = true, width = \textwidth]{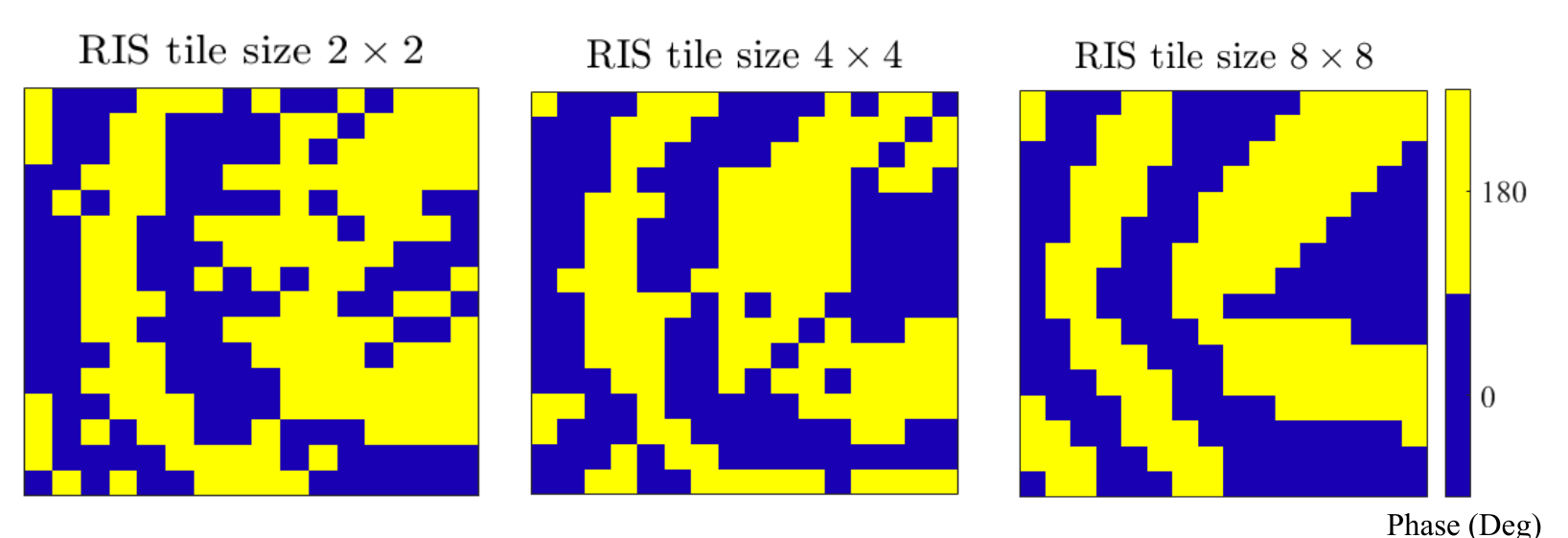}
    \caption{Tx-Rx distance = 2.5 m}
    \label{fig:exppat1.5m}
    \end{subfigure}
    
    \begin{subfigure}[b]{0.5\textwidth}
    \includegraphics[trim = 1cm 7cm 1cm 4.5cm,clip = true, width = \textwidth]{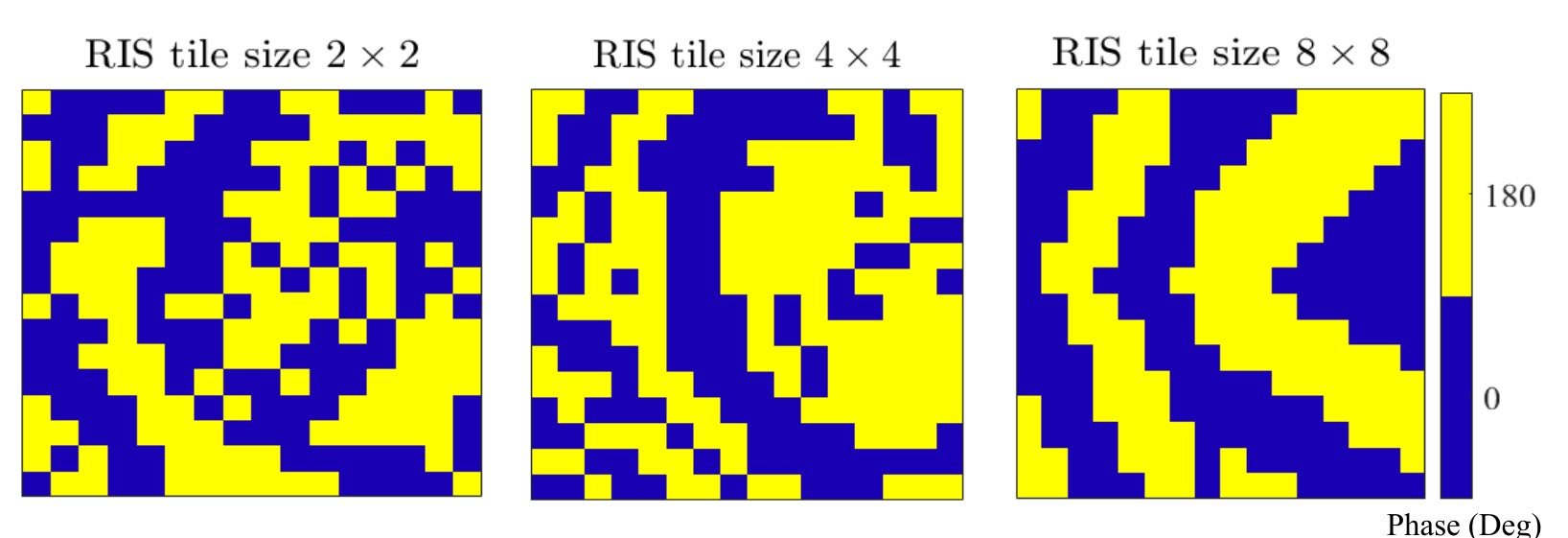}
    \caption{Tx-Rx distance = 3 m}
    \label{fig:exppat2m}
    \end{subfigure}
    \caption{RIS optimal phase distribution according to Tx-Rx distance.}
    \label{fig:exppatvsdist}
\end{figure}

The RIS optimal phase distribution patterns according to different Tx-Rx distances are shown in Fig.~\ref{fig:exppatvsdist}. Convex lens-like patterns are formed for every case. There is a good agreement between the patterns of different RIS tile sizes at each Tx-Rx distance. It is noticed that the patterns with RIS tile size of $2\times2$ and $4\times4$ get worse as the Tx-Rx distance increases (see Fig.~\ref{fig:exppatvsdist}(e)) while that of the RIS tile size of $8\times8$ is fine. This is because the signal reflected from the smaller RIS tile size becomes very small when the receiver moves further away from the RIS. This incurs errors in the calculation of the algorithm. This phenomenon can be observed in Fig.~\ref{fig:Exppoweff}. The receive DC power with RIS tile size of $8\times8$ is higher than those of the RIS tile sizes of $2\times2$ and $4\times4$ at 3 m Tx-Rx distance. It is observed that, in the near region (relatively shorter range from the transmitter), the RIS tile size of $2\times2$ performs the best of all three cases. On the other hand, in the middle region (medium range from the transmitter), the RIS tile size of $4\times4$ outperforms the other cases. The RIS tile size of $8\times8$ is superior to the other cases in the far region (relatively longer range from the transmitter).

\section{Conclusion}
\label{section:conclusion}

In this paper, we have implemented a real-life testbed of RIS-aided WPT system. We have proposed a multi-tile RIS beam scanning algorithm to enable the beam focusing capability of the RIS with only power information. In addition, the mathematical analysis was elaborately presented. Specifically, the RIS tile scanning algorithm was introduced to find the optimal phase and direction control parameters of the RIS tile. Then, multi-tile RIS scanning algorithm was performed by iteratively scanning and optimizing all RIS tiles and the transmitter. The simulation has been carried out to prove the effectiveness of the proposed algorithm. The simulation results showed that the RIS with the proposed MTBS algorithm can greatly improve the receive power at the receiver. 

We have built a real-life testbed of the RIS-aided WPT system and performed the experiment to verify the MTBS algorithm. The experiments with different RIS tile sizes were conducted. The experimented results showed that the proposed algorithm works very well. All RIS tiles were well optimized in every iteration. Especially, in the considered test scenario, an approximately 20 dB gain in the receive power has been observed for all RIS tile sizes in comparison with the ``RIS OFF'' case. Even in the presence of the obstacle, the MTBS algorithm still provided the same performance as the case without the obstacle. It demonstrated that the proposed algorithm provides a great improvement even with the non-light-of-sight (NLOS) channel. The power transfer efficiency has been obtained according to the Tx-Rx distance. The experimented results showed that about 22.5 mW (1.05\% of efficiency) is transferred to the receiver at the Tx-Rx distance of 1.3 m when RIS tile size of $2\times2$ is adopted. We expect that higher transferred power will be achieved by deploying a large-scale RIS.


\bibliographystyle{Bibliography/IEEEtran}
\bibliography{Bibliography/IEEEabrv,Bibliography/IEEE_REF}\ 








\end{document}